\documentclass[a4paper,final]{elsarticle}

\makeatletter
\def\ps@pprintTitle{%
 \let\@oddhead\@empty
 \let\@evenhead\@empty
 \def\@oddfoot{}%
 \let\@evenfoot\@oddfoot}
\makeatother

\usepackage{moreverb}
\usepackage[colorlinks,bookmarksopen,bookmarksnumbered,citecolor=red,urlcolor=red]{hyperref}

\usepackage{xfrac}
\usepackage{graphicx}
\usepackage{listings}
\usepackage[utf8]{inputenc}
\usepackage{color}
\usepackage{xspace}
\usepackage{multirow}
\usepackage{booktabs}
\usepackage{algorithmic}
\usepackage{paralist}
\usepackage{inconsolata}
\usepackage[]{algorithm2e}
\usepackage[]{amsmath}
\usepackage[]{amssymb}
\usepackage{caption}
\usepackage{subcaption}
\usepackage{moresize}

\definecolor{dkgreen}{rgb}{0,0.5,0}

\newcommand{\workingset}{partition\xspace}
\newcommand{\workingsets}{partitions\xspace}
\newcommand{\indpartition}{individual partition\xspace}

\begin{document}

 \begin{frontmatter}

\title{Cache-Conscious   Run-time Decomposition of  Data Parallel Computations}

\tnotetext[label0]{Submitted to Journal of Supercomputing}

\tnotetext[label1]{This work was partially funded by FCT-MEC in the framework of the  PEst-OE/ EEI/UI0527/2014 strategic project.}

\author{Hervé Paulino}
 \ead{herve.paulino@fct.unl.pt}
 \ead[url]{http://asc.di.fct.unl.pt/~herve}

\author{Nuno Delgado}

\address{NOVA Laboratory for Computer Science and Informatics \&  Departamento de Informática, Faculdade de Ciências e Tecnologia, Universidade Nova de Lisboa, 2829-516 Caparica, Portugal}

\begin{abstract}
Multi-core architectures  feature an
intricate hierarchy of cache memories, with multiple  levels and  sizes.
To adequately decompose an application according to the traits of a particular memory hierarchy is a cumbersome task that may be rewarded with significant performance gains. The current state-of-the-art in memory-hierarchy-aware  parallel computing delegates this endeavour on the programmer, demanding from him deep knowledge of both parallel programming and computer  architecture. 
In this paper we propose the shifting of these memory-hierarchy-related concerns to the run-time system, which then takes on the responsibility of distributing the computation's data across the target memory hierarchy.
We evaluate our approach from a performance perspective, comparing it against  the common
 cache-neglectful data decomposition strategy. 
 \end{abstract}

\begin{keyword}
Data parallelism, Data locality, Cache Optimizations, Parallel computing, Run-time support
\end{keyword}

\end{frontmatter}

\section{Introduction}
\label{sec:intro}

Data parallelism is the most common parallel decomposition strategy, by which 
 an 
application's domain is  decomposed  into as many partitions
as workers assigned to the computation.
%Ergo, the domain is partitioned as evenly as possible between these N workers, spawning that many tasks.
%
Such strategy is  cache hierarchy neglectful and hence, in many cases, does not harvest the 
benefits provided by the (consistently growing amount of) cache hardware available in current computers, from laptops to high-end server nodes.

% this neglets the increasing amount of cache -- as hardware only spatial and not temporal

An adequate mapping of a computation onto the underlying memory hierarchy is crucial to fully harness the computational power of modern architectures.
%
%The performance gains derive from either temporal or spatial  locality in the access to data. 
%
%
However,  cache memory management  is completely transparent to user-level programming.
Such responsibility typically falls upon the hardware infrastructure, whose 
function is only to guarantee that recently  accessed data  is closer to the computing unit(s) than the remainder, since it will likely be accessed again.
%
%Accordingly, it handles the issue of spatial locality, but not temporal locality. 
%
Cache replacement algorithms 
are based upon heuristics,  such as \emph{least-recently-used}, 
that   do not always serve the application's best interest, given that  they base their decision
on historic information rather than on information about future accesses.
%
%Once a computation's work-set overflows the capacity of a cache level, 
%the hardware will remove previously fetched data from the cache in order to make room for new data. Although the hardware makes use of heuristics to dedice which data is to be removed, an example being the \emph{least-recently-used} removal strategy, these do not always provide the best removal strategy for particular applications.
%
This limitation has a major impact on the performance of temporal locality sensitive computations, such as stencil  computations and computations over matrices.  

To overcome this problem, it is up to   the compiler, the run-time system or, ultimately, the programmer
to implement efficient, application-specific, mappings that maximize the cache hit ratio.
%
%by having the workload fitting the smallest cache level, one can expect to fully exploit cache locality (including temporal) during the execution of each task.
%
% sota-  cache-consciouss and hierarchical approaches such as ...  -- prog
% the rest
Cache-guided optimizations in mainstream compilers consist essentially of loop transformations \cite{McKinley:1996}  directed
at  sequential  loops,  which, in the context of parallel computing, may only be applied to the internal execution of tasks.
In fact, the data locality issue in parallel computing has been mostly addressed at language level, via 
  linguistic constructions for the explicit programming of the memory  hierarchy  \cite{sequoiapmh,hpt,Biksh06programmingfor,TreichlerBA13,hierarchical_upc,Kamil:EECS-2013}.
However, these  place a heavy burden on the programmer,
requiring in-depth knowledge of parallel programming and computer architecture.
Moreover, they apply divide-and-conquer strategies that cannot be applied in frameworks that cleanly separate the problem decomposition stage from the execution stage, such as MapReduce \cite{mapreduce}. 
The same reasoning may be applied to  cache-oblivious algorithms \cite{cache-oblivious}, which  oblige the programmer to design   
divide-and-conquer algorithms that do not depend on the specificities of a particular cache hierarchy.
%Furthermore, parallelism is not addressed.

In this work we are particularly interested in the aforementioned class of computations, 
where an initial  decomposition stage %(\textit{Split} in MapReduce) 
generates a set of partitions, upon which a computation stage is subsequently applied in parallel.
More precisely, we are  interested on 
exploring how  this decomposition stage may enhance the data locality properties 
of the ensuing computation, so that the latter may  transparently benefit from 
a good mapping onto the
underlying memory hierarchy.  
To carry out such enterprise we  have been researching % \cite{nuno}
on how to 
address some key challenges, namely on
how to deduce mappings for specific  memory hierarchy configurations from the same source code, and on
how to ensure performance portability across a wide range of  configurations.
Overall, the main contributions of this paper are twofold: 
\begin{enumerate}
\item  domain decomposition principles and algorithms that take into consideration 
  the  complex memory hierarchies of current computer architectures, and
   \item  a study that compares our cache-conscious  approach against the classical, cache-neglectful,  strategy.  
\end{enumerate}
To the best of our knowledge, no such comparative study exists. Works such as \cite{sequoiapmh,hpt,Biksh06programmingfor}
simply present speed-up analyses against  sequential versions of the benchmarks.
There is no evidence that the  effort required from the programmer (to explicitly map the computation onto the memory hierarchy) actually delivers performance gains when compared to simpler  strategies. 
%
%how do we evaluate it and the how good are our results
%In order to evaluate our vertical decomposition approach, we compared it against a flat decomposition one; the previously mentioned systems evaluate their approaches by comparing their implementation against sequential ones. 
Conversely, our study attests the validity of our proposal, showing that it delivers  
significant speed-ups to 
computations  that are particularly sensitive to temporal locality.

The remainder of this paper is structured as follows:
Sections \ref{sec:proposal} and   \ref{sec:impl} present the principles and implementation  of  our cache-conscious decomposition of data-parallel computations;
Section \ref{sec:evaluation} evaluates our proposal  from a performance perspective, with a particular focus on the 
comparison against the classical strategy for domain decomposition;
Section \ref{sec:related} positions our approach relatively to  the current state-of-the-art; and, finally, Section \ref{sec:conclusions} presents our final conclusions and prospective future work.

\section{Cache-Conscious Domain Decomposition}
\label{sec:proposal}

Domain decomposition in distributed memory environments is a two stage operation:
initially, the domain is partitioned among the nodes that compose the distributed  system,
and secondly, within each node, it is further partitioned 
 among the worker threads locally assigned  to the computation. 
We argue that these two stages must be clearly decoupled, 
so that stage-specific optimizations   may be devised.
For instance, handling heterogeneity only makes sense at \textit{cluster} level, while cache-awareness only makes sense at \textit{node} level. 

The focus of this  paper is on how to explore  
 domain decomposition (at node level) to enhance the locality properties of data parallel computations.
To that end, we leverage the cache hardware available at each node and apply a cache-conscious strategy that takes into account the 
characteristics  of some (or all) levels of the target memory hierarchy, and not only of the worker threads assigned to the computation (the standard \textit{horizontal} approach).
As a result, the number of resulting \workingsets will be a function of  
 the target machine's cache hierarchy.

\begin{figure}
\centering
\includegraphics[width=0.7\linewidth]{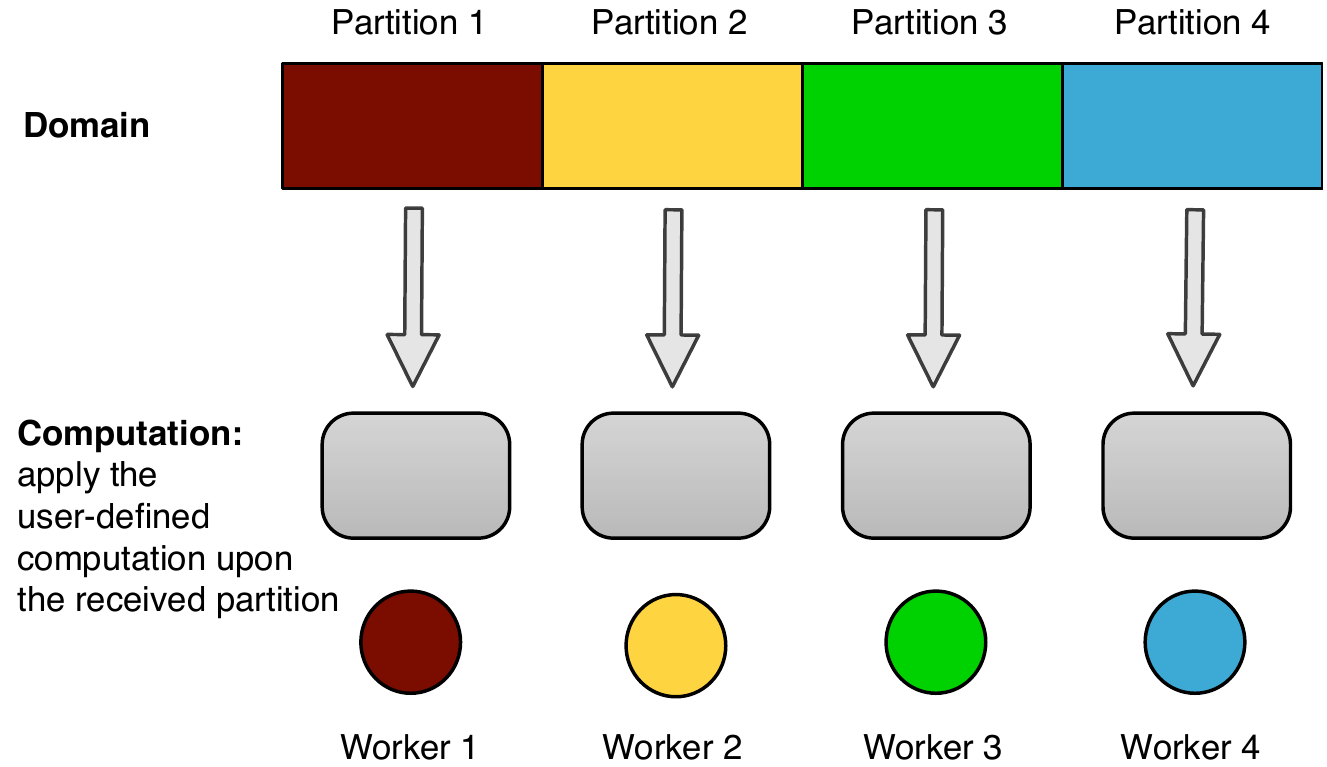}
\caption{Application of a user-defined computation over a horizontally decomposed domain}
\label{fig:horizdecomp}
\end{figure}

\begin{figure}
\centering
\includegraphics[width=0.7\linewidth]{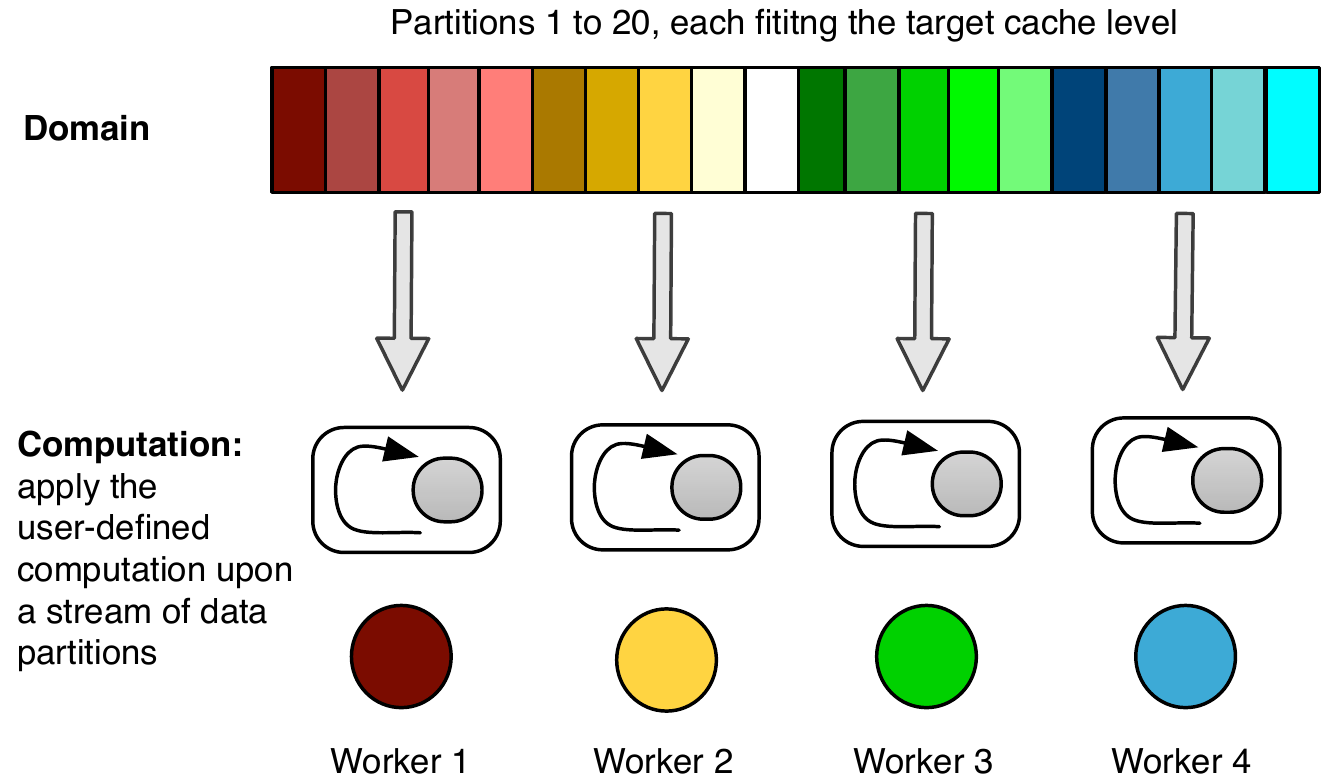}
\caption{Application of a  user-defined computation over a cache-consciously decomposed domain}
\label{fig:ccdecomp}
\end{figure}

Figures \ref{fig:horizdecomp} and  \ref{fig:ccdecomp}  highlight the differences  between  the  horizontal decomposition approach and our  cache-conscious proposal.
In the latter case, the domain is decomposed in such a way that each partition fits  - its size is a function of  - a given target cache level (TCL). Additionally,  the behavior of each worker assigned to the computation is modified so that it iteratively applies   the user-defined computation upon a stream of  partitions, rather than to a single one.
The number of workers is preserved, the amount of work performed by each worker is also generally preserved\footnote{The user-defined decomposition algorithm may generate irregular partitions,  ultimately leading to  an unbalanced  work distribution among the workers. The issue is orthogonal to the cache-conscious decomposition, which in general does not   increase nor decrease such unbalance. }, but the granularity of the data upon which the user-defined computation is individually applied is (potentially) much smaller, thus enhancing locality.


To successfully apply our strategy we have to address two distinct challenges:
i) how to \textit{decompose} a computation's domain so that 
such  computation may make better use of the available cache hardware
 and  ii) how to efficiently
\textit{schedule} the resulting tasks onto 
 the available set of workers,  while also 	leveraging data locality. 

\subsection{Decomposition}
\label{sec:decomp:node}

For the sake of generality, we allow a computation's
 input data-set (our domain to decompose)  to be built from multiple sub-domains, typically individual data-structures, each 
  with its own decomposition strategy.
 For example, in the classic matrix multiplication, one can implement a single decomposition strategy that splits the 3 matrices involved, or decompose the 3 matrices individually and have a way of combining the resulting individual partitions. 
In this latter strategy, a partition of the input data-set  must comprise a partition of each of its sub-domains.

Naturally, the number of indivisible units of each sub-domain may not be a multiple of the number of desired partitions.
This fact may be easily solved by distributing the remainder units among the regular-sized partitions, causing an unbalancing of, at most,  
one indivisible unit.
However, problem-specific constraints may impose further restrictions upon the number of partitions and/or the geometry of the decomposition as a whole. Stencil computations present this kind of restrictions  with regard to the number of elements and  their relative
positions. 
For example, consider 
a computation over  a  2-dimensional grid where, at instant  $t_{i+1}$, the value of each of the grid's elements
is a function of its own value and of its 8 adjacent elements  at instant $t_i$. 
%
% of the 
%
% the computation of each  element in a 2-dimensional  grid at time $t_{i+1}$ as  a function of the values of its position and of its 8 adjacent elements  at time $t_i$. 
%%
To meet these restrictions, each partition of the domain must comprise $3 \times n$ elements, with $n \geq 3$, organized 
in grids   of, at least, $3\ \text{rows} \times 3\ \text{columns}$.

This problem-specific  information must be conveyed in the decomposition algorithms supplied by the programmer, that we refer to as 
\textit{distributions},
 and must therefore be included in the interface that regulates the implementation of such algorithms (Table \ref{lst:dist}).
Additionally, in order to perform our cache-conscious systematic decomposition, 
we will require an extra set of functions, whose purpose we will unveil as we progress along this section.

\begin{table*}
\centering
\begin{scriptsize}
\begin{tabular}{l|p{6cm}}
 partition(\textbf{int} np) : T[] & Partitions the input domain into \texttt{np} partitions \\
 \midrule
 validate(\textbf{int} np) : \textbf{int} & Validates if the input domain may be partitioned into \texttt{np} partitions, result:\\
 &  \quad $< 0$ -  there is no solution for any value $\geq np$, \\
 &  \quad $ = 0$ -  $np$ is not a valid solution, but there may be solutions for values $> np$,  \\
  &  \quad $> 0$ - $np$ is a valid solution. \\
  \midrule
  getElementSize() : \textbf{int} & Size of an element of T (in bytes) \\
  \midrule
getIndivisibleSize(\textbf{int} np) : \textbf{int} & Indivisible size of a partition  (in number of elements) \\
\midrule
  getAveragePartitionSize(\textbf{int} np) : \textbf{float} & Average size of a partition (in number of elements) \\
  \midrule
  getAverageFirstDimSize(\textbf{int} np) : \textbf{float} & Average size of the first dimension of a partition (in number of elements)
\end{tabular}
\end{scriptsize}
\hrule
\caption{The \texttt{Distribution<T>} interface}
\label{lst:dist}
%\vspace{-20pt}
\end{table*}

\subsubsection{Determining the number and size of the partitions}

Given a domain $D$, composed of $d$ sub-domains ($D= \bigcup_{i=0}^{d-1} D_i$),
its decomposition into a set of partitions $P_D$, 
each fitting a given TCL, requires the calculation 
of a value for the number of partitions ($np$) 
such that 
$$\forall p \in P_D , \:  size(TCL) \ge size(p) = \sum_{i=0}^{d-1} \frac{size(D_i)}{np}$$
where $size()$ denotes the size in bytes of either the enclosed cache level or partition.

\begin{algorithm}[t]
\hrule
\begin{small}
\caption{Verification if a domain may be decomposed into a given number of partitions}
\label{alg:iter_partitioning}
\LinesNumbered
\KwIn{\texttt{TCL\_PER\_CORE} - Size in bytes of the TCL per core}
\KwIn{\texttt{CACHE\_LINE\_SIZE} - Size in bytes of a cache coherence line}
\KwIn{$nDomains$ - Number of sub-domains that form the domain to decompose}
\KwIn{$np$ - Number of partitions into which each sub-domain must be decomposed}
\KwIn{$dists$ - Vector holding the distribution algorithms for each sub-domain}
\KwIn{$\varphi$ - Function that estimates the size of a partition in cache}
\KwOut{$1$ - the  value  of $np$ is valid}
 \KwOut{$0$ - the  value  of $np$ is not valid, but higher values may be}
 \KwOut{$-1$ - the  value  of $np$ is not valid, nor are any higher values}
	$totalPartitionSize \leftarrow 0$; \\
%	$valid \leftarrow \textbf{true}$; \\
	 \For{$i \leftarrow 0$ \KwTo $nDomains$}{ 
	    $status \leftarrow dists[i].\texttt{validate}(np)$;\\
	    \lIf{$status <=$ \textbf{\texttt{0}}}{\Return status}
 %    $fDimSize \leftarrow  dists[i].\texttt{getAverageFistDimSize}(np)$;\\
%	 $partSize  \leftarrow \varphi($\texttt{CACHE\_LINE\_SIZE}$, dists[i], np)$;	 \\
		$totalPartitionSize \leftarrow  totalPartitionSize + \varphi($\texttt{CACHE\_LINE\_SIZE}$, dists[i], np)$;  
	}
   	\lIf{$totalPartitionSize \leq \texttt{TCL\_PER\_CORE}$} {\Return 1} 
   	\lElse{\Return 0} 
\end{small}
\hrule
\end{algorithm}

The proposed value of $np$ must be validated by each of the distribution algorithms involved, being
the verification logic  defined in function \texttt{validate}.
Algorithm \ref{alg:iter_partitioning} presents the procedure for  assessing if a given
number of partitions (${np}$) is valid:
it assures that each sub-domain may be split into that many partitions, according to  distribution algorithms involved (line 3), 
and that the cumulative size of such partitions (${totalPartSize}$)  fits in the TCL (line 7).
The algorithm depends on an estimation of how many  bytes a partition will occupy in the TCL.
To enable the experimentation with different heuristics, the estimation   is delegated on function $\varphi$ (line 5), supplied as parameter.
The outcome of the algorithm is the validation, or invalidation, of the candidate $np$ value.
In the later case, information  concerning values higher than the candidate is also supplied.
This information is used to delimit the search space, as subsequently explained.

To compute the optimal size of a \workingset that fits in a TCL, we apply 
a binary  search:
the value of ${np}$ begins in the number of workers assigned to the execution (${nWorkers}$) and doubles in every iteration until 
a valid solution is found or all values larger that $np$ are invalid
(according to Algorithm \ref{alg:iter_partitioning}).
From then  onwards, the search's interval is continuously narrowed  
to find the smallest valid $np$  value. 
Given that the size of each \indpartition is inversely proportional to  ${np}$, such solution is optimal for the provided input parameters.
The ${nWorkers}$ lower-bound guarantees that the algorithm generates, at least, as many  \workingsets as available workers, in order to fully exploit the designated resources.

\subsubsection{The $\varphi$ function}
The definition of the $\varphi$ function  implies a trade-off between accuracy, computational overhead, and wasted cache space.
A simple approach ($\varphi_{s}$) is not to take into consideration either the size of the target architecture's cache line size (\texttt{CACHE\_LINE\_SIZE} in the algorithm) nor the partition's geometry.
Thus, the function simply computes the number of bytes the partition takes: \\

\resizebox{0.9\linewidth}{!}{
\begin{minipage}{\linewidth}
\noindent
\begin{align*}
  & \varphi_s(cacheLineSize, dist,  np)   =    \\
  & \ \ dist.\texttt{getElementSize}()  \times \left \lfloor{dist.\texttt{getAveragePartitionSize}(np)+0.5}\right \rfloor \\
\end{align*}
\end{minipage}
}\\
where the  result of  \texttt{getAveragePartitionSize} is rounded-up to the closest integer to better suit the most common expected partition size.

A more conservative estimate  ($\varphi_{c}$)  considers to some extent the two previously  neglected  dimensions, at the expense of more computational overhead: \\ %Function  \texttt{getAverageFistDimSize} is used to 

\resizebox{\linewidth}{!}{
\begin{minipage}{\linewidth}
\noindent
\begin{align*}
& \varphi_c(cacheLineSize, dist, np)   =  \\
& \ cacheLineSize \times 
\frac{dist.\texttt{getAveragePartitionSize}(np) \times dist.\texttt{getElementSize}()}{dist.\texttt{getAverageFistDimSize}(np)} \times 
(\lceil \frac{dist.\texttt{getAverageFistDimSize}(np)}{cacheLineSize} \rceil  + 1)
\end{align*}\\
\end{minipage}
}

Function \texttt{getAverageFirstDimSize} returns the average length (in number of elements) of the first dimension of the domain.
%\footnote{For performance reasons, functions \texttt{validate} and  \texttt{getAverageFirstDimSize} may be merged into a single function.}.
 %  
This  information is  particularly important for  the decomposition of multi-dimensional domains, specially multi-dimensional arrays\footnote{When targeting other kinds of data-structures, the default return value of \texttt{getAverageFirstDimSize} may simply be 1.}.
We are assuming a row-major order memory layout
%\footnote{The most used approach in current programming languages, including Java (in which we implemented our proposal).}
	and, in such context, 
the output of the \texttt{getAverageFirstDimSize} function is  crucial to understand the breakdown of a partition into cache lines. 
The use of the average value conveys some extra information to the system when the size of the partitions is not uniform, 
as happens when the size in bytes of the sub-domains is not a multiple of $np$.
%, for a given sub-domain $D_i$,  $(size(D_i) \bmod{} np)  > 0$.

In  $\varphi_{c}$, the size of the partition's first dimension is adjusted	to the boundaries of the cache line.
Furthermore, an extra cache line is added to consider the eventual misalignment of the partition to such boundaries.
This approach is  likely to ensure that the entire working set fits the TCL, but its conservative nature 
will eventually waste more  space than the first approach.
Table \ref{tab:app} illustrates, for both approaches,  the estimated number of bytes that a partition of size $size(p)$ 
will take in a cache with  cache line size $size(cl)$.

\begin{figure}
	\centering
	\includegraphics[width=10cm]{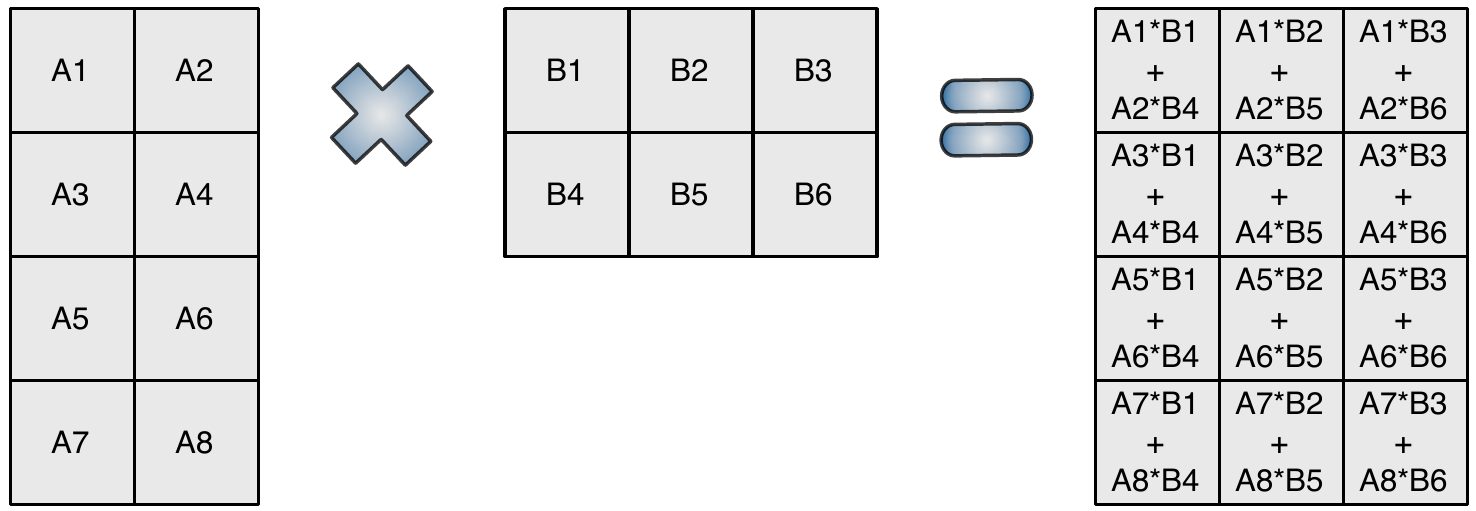}
	\caption{Block decomposition for the matrix multiplication problem}
	\label{fig:matrix_combs}
\end{figure}

As an example, consider  the block decomposition for the parallel computation of the 
classic matrix multiplication problem illustrated in  Figure \ref{fig:matrix_combs}.
A  partition of the domain must comprise a block of each input matrix and space for the computed result block, to be placed in the output matrix.
Note, however, that every block partition of the first matrix ($A$) must be paired with all the block partitions that compose a line of the second ($B$).
%In the  example of  Figure \ref{fig:matrix_combs}, 
%each block partition of matrix $A$ is paired with three block partitions  of matrix $B$, for instance block $A_1$  with blocks $B_1$ to $B_3$.
%
Consider now a concrete instance of the problem where both $A$ and $B$ are $1024 \times 1024$ square matrices of 4-byte-long integers, and a  TCL with 64 KBytes.
For tentative  $np$ value   $256 = 16 \times 16$%\footnote{256 is the value obtained when using the matrix distribution algorithm that we have implemented (Listing \ref{lst:distexample}). Other  algorithms may produce  different results.}.  
 the estimation given by  $\varphi_{s}$  is $(\frac{1024}{16})^2 \times 3\ matrices \times 4\ bytes = 49152\ bytes$, 
while the one given by  $\varphi_{c}$ is $64 \times \frac{(\sfrac{1024}{16})^2   \times 3\ matrices \times 4\ bytes}{\sfrac{1024}{16}} 
\times (\lceil \frac{\sfrac{1024}{16}}{64} \rceil + 1) = 64 \times 64  \times 3\ matrices \times 4\ bytes \times (1 + 1) =  98304\  bytes$.
Thus $np = 256$  is valid when using $\varphi_{s}$ but not when using $\varphi_{c}$. % (not fitting the TCL).
%

%
%the \workingsets resulting from the  cache-conscious decomposition of this input data-set (comprising three matrices)
%will feature three blocks whose cumulative size totals 65368 bytes. 

\begin{table*}
\centering
\resizebox{\textwidth}{!}{
\begin{tabular}{|c|l|c|c||c|c|}
\cline{3-6}
\multicolumn{2}{c|}{}  & \multicolumn{2}{c||}{$\varphi_{s}$}  & \multicolumn{2}{c|}{$\varphi_{c}$} \\
\cline{3-6}
\cline{3-6}
%\multicolumn{2}{c|}{}  & \multicolumn{2}{c||}{Boundary-aligned}   & \multicolumn{2}{c|}{Boundary-aligned}  \\
%\cline{3-6}
\multicolumn{2}{c|}{} &	 Boundary-aligned & Not boundary-aligned &	 Boundary-aligned & Not boundary-aligned \\
\hline
\multirow{2}{2.2cm}{Multiple of cache line size} & Yes& $size(p)$ & $size(p)$ & $size(p) + size(cl)$ & $size(p) + size(cl)$\\
\cline{2-6}
& No & $size(p)$ & $size(p)$ & $size(cl) \times \frac{size(p)}{F} \times (\lceil \frac{F}{size(cl)} \rceil  + 1)$
& $size(cl) \times \frac{size(p)}{F} \times (\lceil \frac{F}{size(cl)} \rceil  + 1)$\\
%& No & $size(p)$ & $size(p)$ & $size(cl) \times ( \lceil \frac{size(p)}
 % {F \times size(cl)} \rceil  + 1)$
%& $size(cl) \times ( \lceil \frac{size(p)}
%  {F \times size(cl)} \rceil  + 1)$\\
\hline
\end{tabular}
}
%\vspace{-5pt}
\caption{Estimated number of bytes  that a multi-dimensional partition of size $size(p)$ bytes (whose first dimension comprises $F$ bytes)
 occupies in a cache with line size $size(cl)$ bytes.}
\label{tab:app}
\end{table*}

None of the presented $\varphi$ functions take into consideration  set associativity.
The actual location of the data in the process' addressing space is not made available in 
 many programming languages, e.g. Java.  
Moreover, the computational complexity  required to take such knowledge into consideration would, most likely, subsume the eventual benefits.
Nonetheless, considering a cache replacement algorithm of the \emph{least-recently-used} family,
 the subjugation of a \workingset's size  to the TCL's capacity 
 highly contributes for having the delimited data fully loaded (minus conflicting cache lines) on such cache level. 
% 

%....  scenarios are the 
%In $n$-way associative cache when the \workingset is not composed of totally contiguous data. 
%
%Extreme

%\vspace{-5pt}
\subsection{Scheduling}

\label{sec:scheduling}

The scheduling stage   assigns  pairs (instance of the computation,  associated \workingset) - our \textit{tasks} - to a set of workers.
The problem diverges from the common scheduling of data-parallel computations because the amount of tasks generated by the cache-conscious  decomposition approach largely exceeds the number of
cores  available in the machine.
%For instance, a block decomposition for the parallel computation of the  matrix multiplication problem
%applied to three 1024x1024 matrices, produces 2744 tasks for  a   64KBytes TCL.
%%

%A  partition of the domain must comprise a block of each input matrix and space for the computed result block, to be placed in the output matrix.
%Note, however, that every block partition of the first matrix ($A$) must be paired with all the block partitions that compose a line of the second ($B$).
%%In the  example of  Figure \ref{fig:matrix_combs}, 
%%each block partition of matrix $A$ is paired with three block partitions  of matrix $B$, for instance block $A_1$  with blocks $B_1$ to $B_3$.
%%
%Consider now a concrete instance of the problem where both $A$ and $B$ are $1024 \times 1024$ square matrices of 4-byte-long integers, and a  TCL with 64 KBytes: 
%the \workingsets resulting from the  cache-conscious decomposition of this input data-set (comprising three matrices)
%will feature three blocks whose cumulative size totals 65368 bytes. 
%%
%In order to produce blocks with this size, each matrix will be divided into $14 \times 14 = 196$ blocks. 
%Since each block of $A$ will have to be combined with $14$	 blocks of $B$, applying a one-to-one mapping from \workingsets to tasks will 
%result in a total of $14 \times 14 \times 14 = 2744$ tasks.

Revisiting the application of the
 matrix multiplication algorithm of Figure \ref{fig:matrix_combs} to matrices of dimension $1024 \times 1024$,
 each block of matrix $A$ will have to be combined with $16$	 blocks of matrix $B$.
 Applying a one-to-one mapping from \workingsets to tasks will 
result in a total of $16 \times 16 \times 16 = 4096$ tasks.

Spawning as many workers as tasks is not viable in this context, as 
having the number of execution flows far exceeding the number of computing resources penalizes performance.
Also, given the small granularity of each task, to have a pure dynamic 
work-stealing-based scheduling policy will lead to considerable overheads, since the worker threads will spend a non-negligible percentage of their time fetching work, rather than executing it. 
Thus, performance in this context is highly dependent of an efficient and locality-aware mapping
between tasks and workers.
Our solution is to  perform an initial static scheduling that assigns a cluster of tasks to each worker, which
% executes a coarser grained task that  
 sequentially iterates upon the stream  of the said tasks (recall Figure \ref{fig:ccdecomp}).
%
%This static work distribution increases the system's overall performance.
We advocate that this static work distribution increases the system's overall performance, since
workers do not have to search and compete for work.
Nonetheless, when in the presence of irregular computations,  dynamic scheduling techniques may also be useful to balance the load across the workers.
We do  not yet explore such techniques,  even though we are aware of work in the field  that already embeds some hierarchical concerns \cite{hws,charm}.

%Since we only studied applications featuring regular parallelism and our scheduling strategies attempt to assign a number of tasks as even as possible to each worker, static scheduling makes the most sense since workers are expected to finish executing their set of tasks at nearly the same time.

In the scope of this work we present two distinct task clustering strategies. 
The challenge is to trade-off the schedule's efficiency  against the overhead (temporal and spatial) that the determination of the next task to execute  might impose on the overall execution. More complex clustering strategies are likely to perform more calculations and require more memory, thus stealing space in the cache for the actual computation's data.
%

%It is worth noting that since the task scheduling is done statically, the access to the global task set can be free of any synchronization mechanism, avoiding any overhead that could result from synchronization mechanisms.

\subsubsection{Contiguous Clustering (CC)}
\label{subsec:cc_clustering}
\begin{figure}[t!]
	\centering
	\includegraphics[width=.9\linewidth]{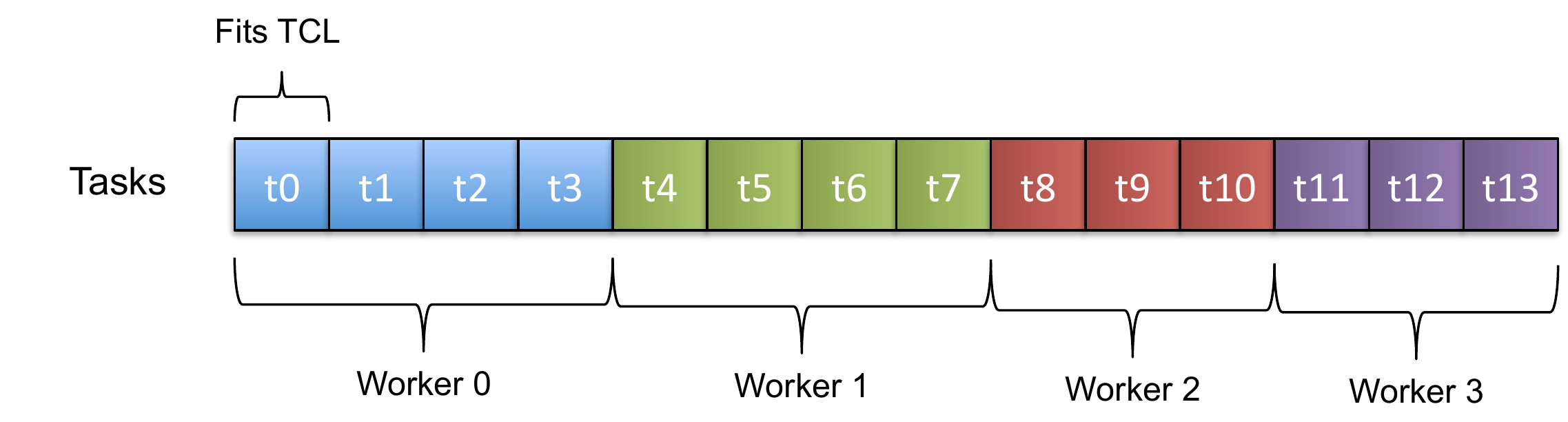}
	\caption{Contiguous Clustering: scheduling of 14 tasks among 4 workers}
	\label{fig:cc_clustering}
\end{figure}

Assigns an equally-sized contiguous cluster of tasks to each worker, according to their unique identifier. 
Given a set of $n$ workers and a set of $m$ tasks,  a worker with rank $i$ is 
assigned the following cluster of tasks: $[(i\times \frac{m}{n}), ((i+1)\times \frac{m}{n})[$.
Whenever  $m$  is not a multiple of $n$, the first $r$ workers (for  $r = m \bmod{} n$)  receive one extra task. 
Figure \ref{fig:cc_clustering} illustrates the application of the CC strategy to the scheduling   
of 14 tasks among 4 workers.

The rationale behind this strategy is twofold: 1) introduce minimal overhead during scheduling, and 2) exploit spatial locality between tasks
operating upon consecutive partitions.
% executions. % when consecutive tasks operate over contiguous data. 
%
%We have observed that it is  very common for 
%two consecutive partitions to comprise contiguous data.  

%Naturally, there are execptions, such 
%	The distributions we implemented for the studied problems guarantee that, whenever possible (line transitions in matrices pose exceptions), sequentially ordered partitions of a given domain contain contiguous data.
% \textbf{dizer que as distribuições que temos/implementámos assim o fazem?}

\subsubsection{Sibling Round-Robin Clustering (SRRC)}
Builds on the fact that the  Last Level Cache (LLC) is shared by multiple computing cores.
Thus, if two or more workers running in such cores share data, the number of LLC misses will decrement, reducing the accesses to main memory. 
Once again, a  matrix multiplication example is paradigmatic, as multiple  partitions share blocks of both input matrices.

%The set of workers $W$ can be obtained by the union of the sets of cores sharing a LLC, hence $W = \bigcup \limits_{i=0}^{n-1} Q_i$, where each set $Q_i$ represents a set of cores sharing a LLC. We will assume that the cardinality of every subset $W_i \subset W$ is the same.

The scheduling algorithm comprises two distinct assignment levels: the \emph{cluster-assignment} level assigns clusters of tasks to groups of workers, whilst the \emph{task-assignment}  level assigns tasks within a cluster to workers within a worker group. 
In the first level,  the size of the task clusters
is ruled by the TCL to LLC ratio: 

\resizebox{\linewidth}{!}{
\begin{minipage}{\linewidth}
\noindent
\begin{align*}
clusterSize = \frac{size({LLC})}{size({TCL})} + (cores({LLC}) - (\frac{size({LLC})}{size({TCL})} \bmod{}  cores({LLC}))) \\
\end{align*}
\end{minipage}
}
The second term simply ensures a proper distribution of the work
when in the presence of remainder.
It uses  the  number of cores that share a LLC, denoted by $cores({LLC})$, as the distribution unit.

\begin{figure}
\centering
\includegraphics[width=0.9\linewidth]{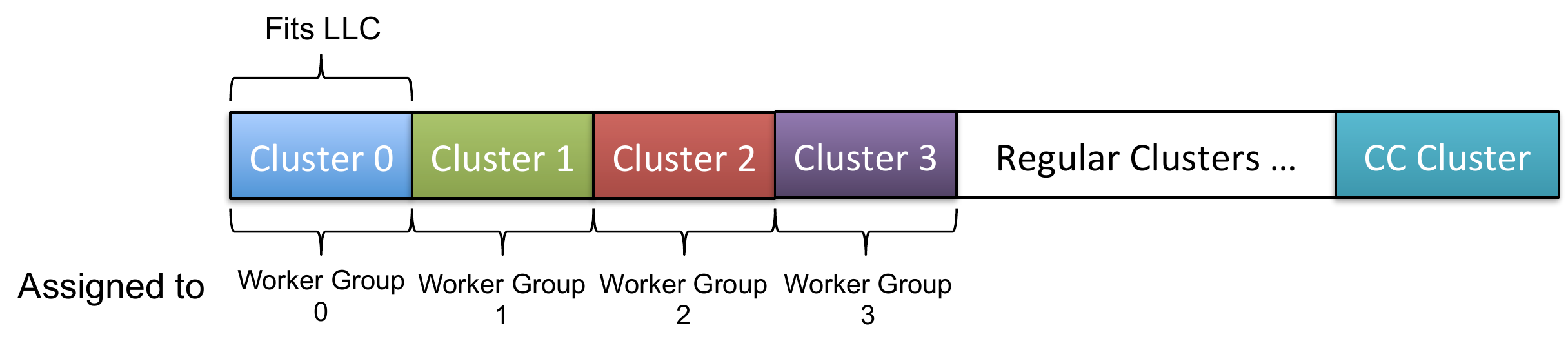}
	\caption{SRRC cluster-assignment: assignment of clusters of tasks among 4 groups of workers running on sibling cores that share a LLC.}
	\label{fig:srr_clusters}
	\includegraphics[width=0.9\linewidth]{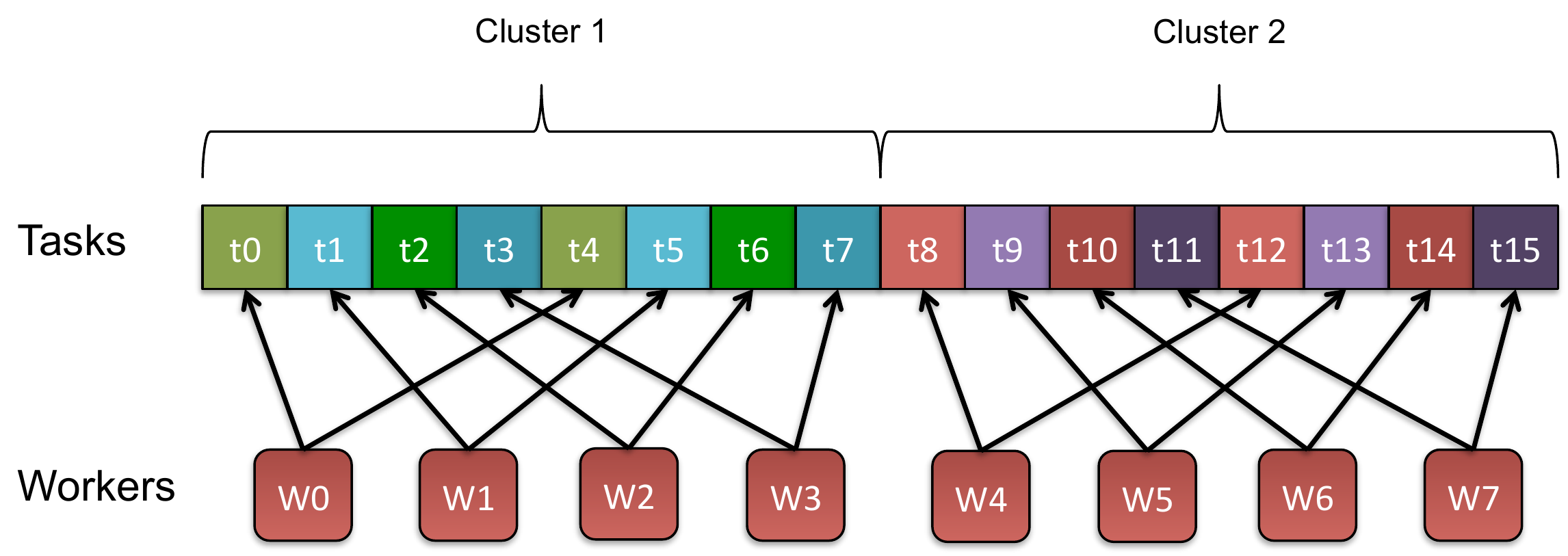}
	\caption{SRRC task-assignment: assignment of the tasks within a  cluster among the workers that compose the target group. }
	\label{fig:srr_clustering}
\end{figure}

Let we represent the resulting set of clusters ($C$) as follows $C = \{ c_0, ...,  c_{n_c-1}\} $, where $c_i$ denotes a particular cluster, and $n_c$ the number of clusters.
Consider now  that the  set of  workers ($W$) may itself be grouped according to 
affinities of the cores (where they will carry out their execution) to the LLC. % (see Section \ref{sec:scheduling:affinities}).
Accordingly, $W$ may be represented as $W = \{ w_0, ...,  w_{n_w-1}\}$ where $w_i$ denotes a particular worker-group and
$n_w$ denotes the  number
of such groups.
Given these definitions, the scheduling of $C$  among $W$
follows a round-robin strategy that
assigns, to each group of workers $w_i \in W$,
the following subset of  clusters: 
$$C_i = \{c_j \in C | j \bmod{} n_w = i \wedge j < (n_c - (n_c \bmod{} n_w))\}$$

To guarantee a schedule as even as possible, when the number of clusters is not a multiple of work groups,
the remainder clusters are merged in a special cluster, named \emph{CC Cluster}.
This cluster  also comprises the  tasks that could not form a cluster (given by $clusterSize \times (n_t \bmod{} (clusterSize \times n_w))$) and is scheduled   according to the  CC  strategy.

Figure \ref{fig:srr_clusters} illustrates 
the cluster-assignment for a machine with 4 groups of sibling cores sharing 4 different LLCs, 
along with the assignment of the resulting clusters to the worker groups. 
The task-assignment within a cluster (illustrated in Figure \ref{fig:srr_clustering})   is performed in a \emph{round-robin} fashion among the	
workers   that compose the group.
%Each worker $w_i$ on the worker group $W' = \{w_0,w_1,...,w_{n-1}\}$ that was assigned the cluster of tasks $T = \{t_0,t_1,...,t_{k-1}\}$ will execute the subset of $T$ containing all tasks $t_j$ such that $j \bmod{} n = i$.
%
%

\subsection{Worker-Core Affinity}
\label{sec:scheduling:affinities}
The thread scheduling policies of modern operating system are aware of the threads' cache footprints, keeping threads on the same core as much as possible.
Nonetheless, in some cases we need to compulsorily constrict the set of cores on which a thread may execute, namely
to properly apply the SRRC strategy.
This strategy assumes that the workers operating over a given task cluster execute on cores sharing a LLC. 
Therefore, it is important to map the affinity between workers and cores accordingly.

In order not to be too restrictive, we allow worker threads to be freely scheduled among the cores
under the lowest shared cache level, a strategy we have adequately baptized as \emph{Lowest Level Shared Cache} affinity mapping. 
As an example, in a quad-core architecture 
with dedicated L1 caches, a L2 cache shared by each two cores, and a single L3 cache,
the Lowest-Level-Shared-Cache affinity mapping  allows the operating system  
to freely schedule worker threads between every two cores that share a L2 cache.

%\vspace{-5pt}
\subsection{Synchronization-free Execution Engine}
\label{subsec:hierarchyaware}
To leverage the devised cache-conscious domain decomposition strategy, the underlying execution engine must 
provide efficient access from each worker thread to the tasks assigned to it  by the scheduling policy in place.
 Our approach is to allow the workers to  directly access  
  the tasks generated by   
decomposition stage, which are stored contiguously in a  vector.
This  avoids the performance penalty of moving   tasks  to worker-local data-structures.
Accordingly, each thread iterates through the shared vector to  sequentially fetch and execute each task scheduled to it.

All accesses to the shared task vector are synchronization-free.
This is possible because each worker receives a disjoint set of task clusters, and  the associated 
index sets are locally computable.
From its  unique rank identifier, a worker is able to  
determine the index of the first task to execute, and the relative position of all the remainder.
The computational complexity of this operation is bound to the scheduling policy.
%It is quite more complex in the SRRC approach that in the CC counterpart (where it is trivial).
%
The SRRC approach requires two loops to iterate over the whole set of tasks assigned to it
and needs to deal with  remainders within  and across clusters. In turn, the CC counterpart requires a single  loop over a contiguous vector.

\section{Implementation Details}
\label{sec:impl}

We have prototyped our proposal in the Elina Java parallel  computing framework \cite{elina,DBLP:journals/jcss/PaulinoM15}
for  distributed and shared-memory environments.
Elina supports both embarrassingly parallel data-parallel  and MapReduce computations.
Moreover, it is a very modular framework that cleanly separates most of the system-level functionalities into independent modules,
whose concrete implementation is specified via a configuration file
As a result, equipping the framework with a new implementation of given module,  such as the decomposition or scheduling strategy,
requires \textit{only} the implementation of a pre-defined interface, and
an altering a configuration file.
This option allowed us to, on one hand, have  programming and execution models close to what may be found in the most used
MapReduce-based frameworks for the processing of large data-sets, such as Hadoop \cite{hadoop}, and on the other hand, 
 to evaluate multiple system solutions by simply modifying the framework's configuration, without having to 
modify  the  framework's core or the application's source code.

Our implementation efforts were thus directed to the development of multiple  adapter modules, namely for the domain decomposition strategy (cache-conscious and horizontal), the scheduling algorithm (\emph{Contiguous Clustering} and \emph{Sibling Round-Robin Clustering}), the \textit{Lowest Level Shared Cache} affinity mapping, among others.
Here we will just briefly review how we uniformly represent memory hierarchies in the framework, and how we have implemented the thread-to-cache affinities. 

\subsection{Platform-Independent Representation of a Memory Hierarchy}
There is no standard format for the storing of 
hardware related information by operating systems, nor there is a  single tool that provides such information for (at least) 
the most known operating systems, such as Windows, Linux and Mac OS X.
Accordingly, we had to develop our own platform-independent representation of a memory hierarchy.
For that purpose we use the JSON data representation format.
A memory hierarchy representation is hence defined as a set of  
nested JSON objects comprising the following fields:
\begin{itemize}
\item[\texttt{size} -]  size of each individual memory element  that composes the current memory level (in bytes), such as the RAM memory associated to a given CPU or the size of a particular cache memory.
\item [\texttt{cacheLineSize} -]  size of the cache coherency line (in bytes). This field is present only if the memory level represents a cache level.
\item [\texttt{siblings} -]  array of arrays of sibling cores sharing each copy of the memory level.
\item [\texttt{child} -]  object containing the memory level information of the lower (child) level (is null if the current level is the bottom-most in the hierarchy).
\end{itemize}

As a proof of concept we  implemented a tool that  automatically generates a platform-independent JSON representation of a node's memory hierarchy from 
 the information stored in the 
\texttt{/sys/devices/system/cpu/} directory of a Linux installation.
Listing \ref{lst:json_hier1} showcases the result obtained for a  NUMA (Non-Uniform Memory Access) architecture comprising two quad-core CPUs and 8 GBytes of RAM. 
%Each CPU is bound to 4 Gbytes of local memory, and features a cache hierarchy composed of a single L3, and four L1 and L2 caches.

%
\begin{lstlisting}[float,caption=A NUMA node comprising two quad-core CPUs and 8 GBytes of RAM - 4 Gbytes per CPU. Each CPU features a single L3 cache and four  L1 and L2 caches - one per core., label=lst:json_hier1,language=java]
{
  "siblings": [[0,2,4,6],[1,3,5,7]],
  "size": 4294967296,
  "child": {
    "siblings": [[0,2,4,6],[1,3,5,7]],
    "size": 6291456,
    "cacheLineSize": 64,
    "child": {
      "siblings": [[0],[1],[2],[3],[4],[5],[6],[7]],
      "size": 524288,
      "cacheLineSize": 64,
      "child": {
        "siblings": [[0],[1],[2],[3],[4],[5],[6],[7]],
        "size": 65536,
        "cacheLineSize": 64,
        "child": null
      }
    }
  }
}
\end{lstlisting}

\subsection{Thread to Cache Affinities in Java}
The Java standard API does not provide means for 
establishing affinities between threads running in the Java virtual machine and cores of the underlying processor(s).
Ergo, to implement the \textit{Lowest Level Shared Cache} affinity mapping
we had to resort to external commands, namely 
\texttt{jstack} to obtain 
the correspondence between the Java virtual machine's and the operating system's thread identification, 
and 
\texttt{taskset} to set the affinity of the threads according to the \textit{Lowest Level Shared Cache} policy. 
Note that while the \texttt{jstack} tool is cross-platform,  \texttt{taskset}  is only available in (most) Linux distributions.

\section{Evaluation}
\label{sec:evaluation}

Our experimental evaluation aims to characterize which kind of applications may benefit from the devised cache-conscious decomposition approach, and to quantify such benefit.
For that purpose, we carried out a comparative performance analysis 
against the pre-existing horizontal work distribution featured in Elina.
Elina's portable programming model and run-time system allowed us to use the  same source code on both settings.
We simply deployed the run-time system with different  instances of several modules, namely
of the decomposition strategy,  the scheduling strategy, and  the $\varphi$ function.

\subsection{Test Infrastructure}
All experiments were conducted on two different types of nodes, with the following characteristics:
%on a 8-node (64 hardware threads) asymmetric cluster infrastructure comprising 5 nodes of type A and 3 of type I, with the following characteristics:
\begin{description}%[\labelwidth=1.6.5cm]
%node1
\item[\textbf{System A} -]  2 Quad-Core AMD Opteron\texttrademark\ Processor 2376 with three cache levels: a  64KBytes L1  data cache per core, a  unified 512KBytes L2 cache per core, and a unified 6MBytes L3 cache per processor.
%
%node8
\item[\textbf{System I} -] 2 Dual-Core Intel(R) Xeon(R) CPU X30 hyperthreaded with three cache levels: a 32KBytes L1 data cache  per core, a  unified 256KBytes L2 cache per core, and a unified 8MBytes L3 cache per processor.
%
%node10
%	\textit{System 3 (S3)} - [64-core node] 4 16-Core AMD Opteron\texttrademark\ Processor 6272 with three cache levels:  16KBytes L1 data cache per core; a unified 2MBytes L2 cache per two cores, and one unified 6MBytes L3 caches per eight cores.
\end{description}
All nodes run the Debian Linux distribution with Linux kernel version 2.6.26-2-amd64.  The installed Java platform is OpenJDK 7 (version 1.7.0\_21).
%Although the system is heterogeneous, the relative performance of the nodes is quite similar,  when all 8 hardware threads are in use.
%
%All experiments conducted in this study resort to such configuration, and hence
%the heterogeneity issue may be overlooked.

\subsection{Benchmarks}
To conduct our study we chose
seven benchmarks.
The first two are 
widely used operations over matrices, namely matrix multiplication (\textbf{MatMult})
and matrix transpose (\textbf{MatTrans}). 
The problem  class indicator for both benchmarks represents the dimension of the matrices involved (only square matrices were considered).

Gaussian Blur (\textbf{GaussianBlur}) blurs an image (represented as a matrix) by convolving it with a Gaussian function. 
The benchmark  requires two parameters: the image to blur and the radius of the blurring window for each pixel.
The problem class indicator follows notation $S$-$R$, where $S$ stands for the target image's dimension and $R$ 
for  the blurring radius.

\textbf{Crypt}, \textbf{SOR} and \textbf{Series} are adapted  from the JavaGrande benchmark suite.
Crypt was adapted to cipher and decipher  files instead of messages.
The problem class indicator is bound  to the size of the  file to cipher/decipher.
SOR computes the Successive Over Relaxation algorithm for a matrix of dimension $N \times N$, whilst
Series computes the first $N$ Fourier coefficients of the function $f(x) = (x+1)^x$ on the interval [0,2].
On both these benchmarks,  the problem class indicator denotes the value of parameter $N$.

Finally, \textbf{WordCount} is an implementation of the classic word-count MapReduce example.
As in Crypt, the  problem class indicator is  bound to the size of the  file to process.

\subsection{Implementing the \texttt{Distribution} Interface}
We argue that the effort that our approach demands from the programmer is relatively small, when compared 
with the potential performance gains -  a claim that cannot be sustained by systems that promote the  
explicit programming of the memory hierarchy, such as Sequoia \cite{sequoiapmh}, due to the programming labour involved. 

\begin{scriptsize}
\begin{lstlisting}[language=java, float, caption=Distribution algorithm for  the cache-conscious decomposition of two-dimensional arrays,label=lst:distexample,numbers=left]
public class IntArray2DDistribution extends AbstractDistribution<int[]> {

   private final int numColumns;
   private final int numRows;

   ... // constructors
  
  public int[][] partition(int np) {
    // partition a matrix into np blocks
  }

  public int validate(int np) {
    float sqrt = Math.sqrt(np);  
    float rsqrt = Math.round(sqrt);  // may be cached for performance reasons
    return (sqrt == rsqrt) ? 1 : 0;
  }
    
  public int getIndivisibleSize(int np) {
    return 1;
  }
   
  public float getAveragePartitionSize(int np) {
    float rsqrt = Math.round(Math.sqrt(np)); // may read value from cache
    return (numColumns * numRows)/(rsqrt*rsqrt);
  }

  public float getAverageFirstDimSize(int np) {
    float rsqrt = Math.round(Math.sqrt(np)); // may read value from cache
    return numColumns/rsqrt; 
  }
}
\end{lstlisting}
\end{scriptsize}

To justify our allegation, we present 
in Listing \ref{lst:distexample}
a concrete  implementation of the \texttt{Distribution} interface  for the decomposition of two-dimensional integer arrays.
The focus is on the modifications imposed on the pre-existent 
distribution algorithm, in order for it to comply to the new interface\footnote{This compliance is a necessary and sufficient condition for the application of the  cache-conscious decomposition strategy.}.
For that reason we omit the implementation of method \texttt{partition}, given that it is independent of the decomposition strategy.
Method \texttt{getElementSize} is also absent but only  for code reuse reasons.
The implementation is inherited from the  base class, and simply returns the size (in bytes) of an element of the array: 4 in this case.
The implementation of the remainder methods is fully depicted in the listing. 

Note that the distribution forces  the array to be partitioned into as many blocks per column 
as for row, hence why the \texttt{validate} method forces $np$ to be a perfect square.

%\vspace{-5pt}
\subsection{Performance Evaluation}

\subsubsection{Cache-conscious versus horizontal decomposition} 	
Tables \ref{tab:node-results} and \ref{tab:node-results2} depict the performance gains relatively to the canonical 
horizontal decomposition and a sequential version of the benchmarks.
A color scale allows to quickly identify the best and worst case scenarios.
Darker the background color,  better is the result.
The comparison with the sequential version of the benchmarks are presented only to 
a given a
better perceptive of the overall gains delivered by either approach.

\begin{table}
\centering
\includegraphics[width=0.8\linewidth]{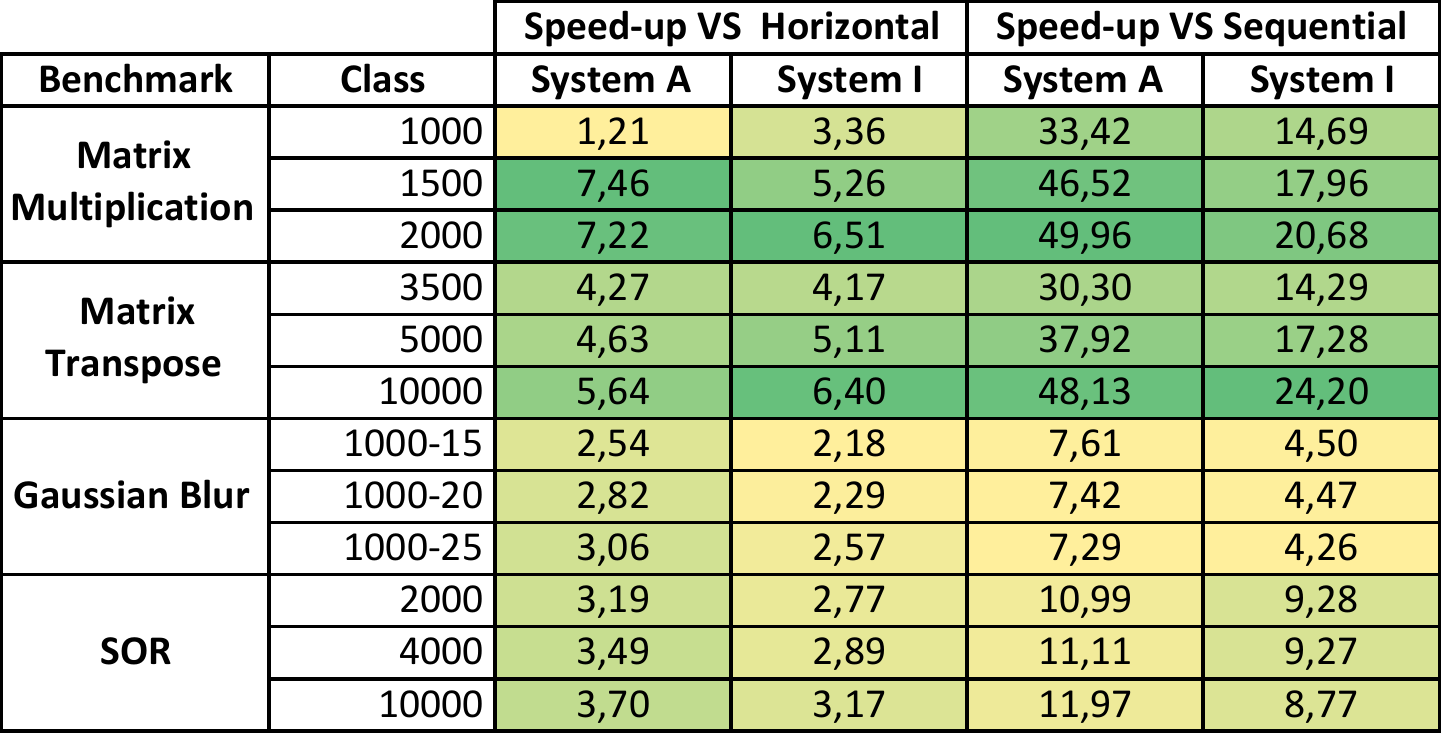}
\caption{Cache-conscious  versus horizontal decomposition and versus the sequential version - Temporal locality sensitive benchmarks}
\label{tab:node-results} 
\includegraphics[width=0.8\linewidth]{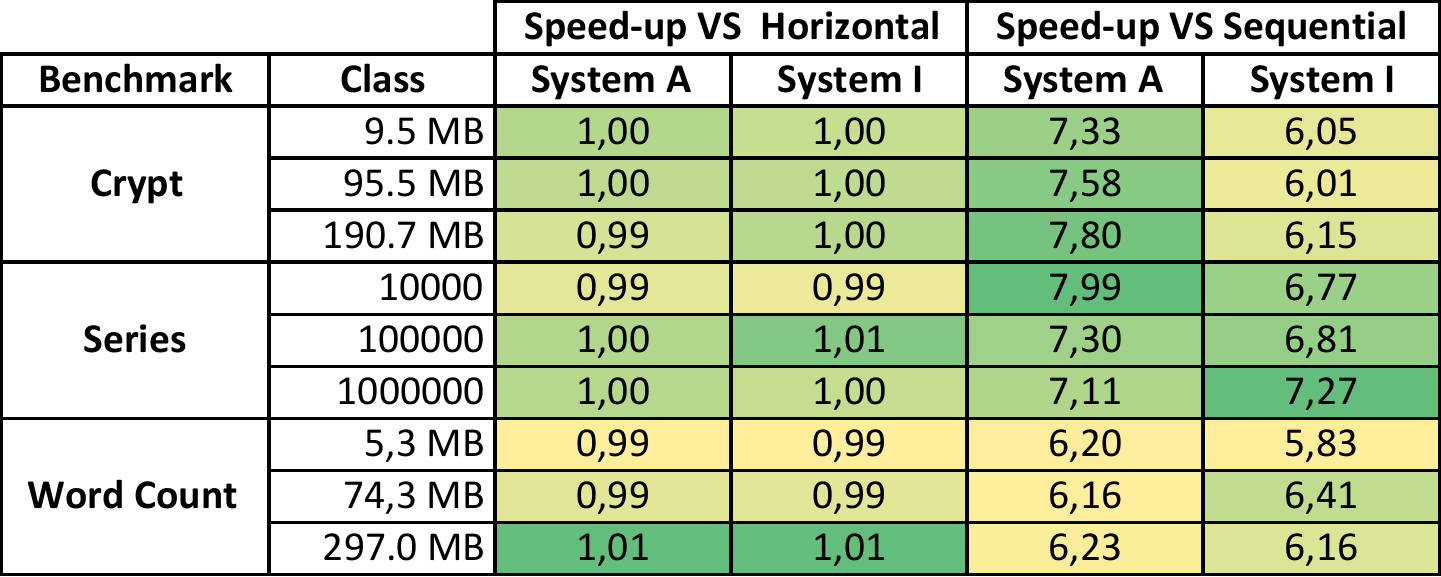}
\caption{Cache-conscious  versus horizontal decomposition and versus the sequential version - No temporal locality sensitive benchmarks}
\label{tab:node-results2} 
%\vspace{-10pt}
\end{table}

\begin{figure}
\centering
\includegraphics[width=0.8\linewidth]{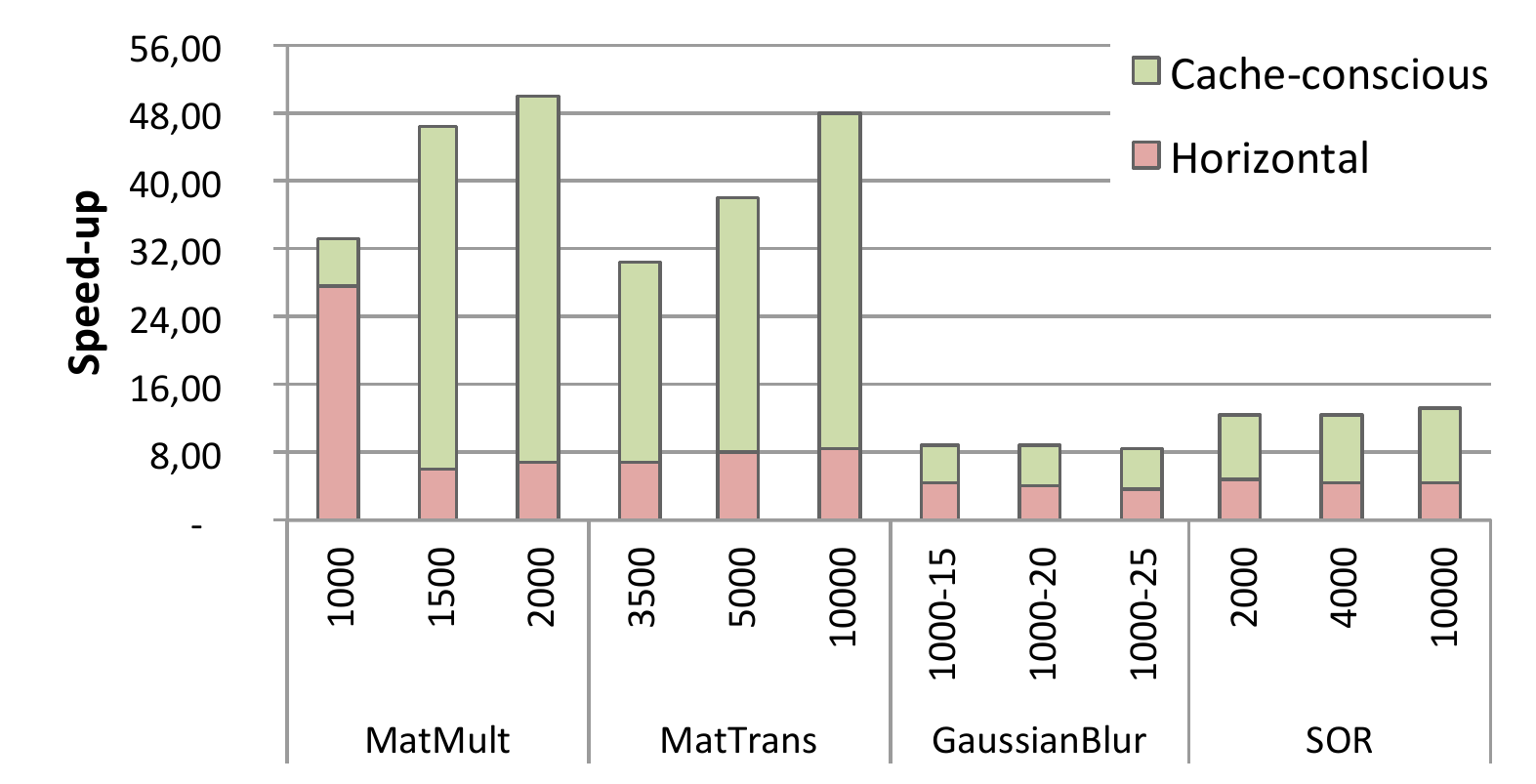}
\caption{Speed-up of both decomposition approaches vs the sequential implementation in System A}
\label{fig:speedupA}
\includegraphics[width=0.8\linewidth]{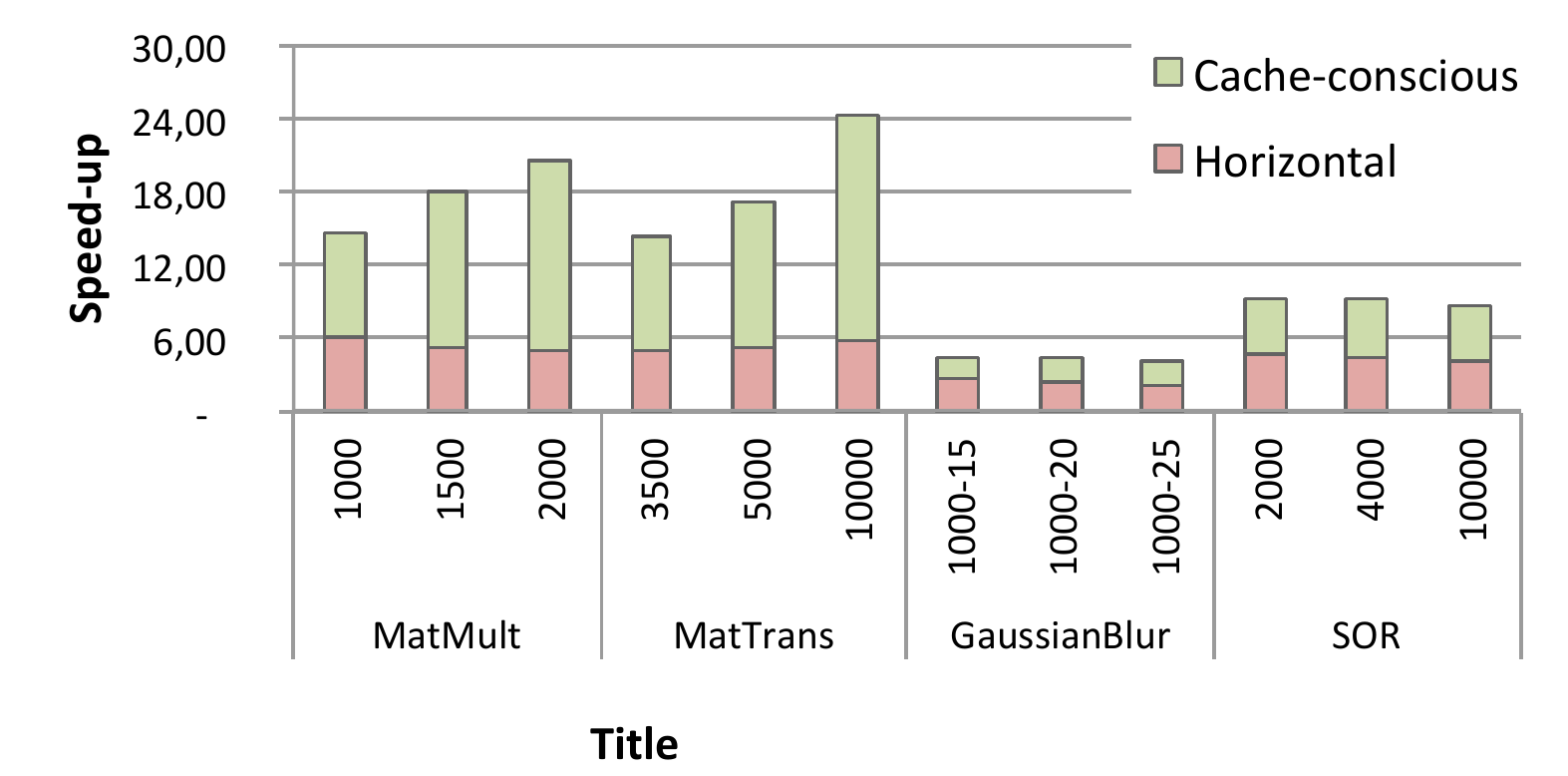}
\caption{Speed-up of both decomposition approaches vs the sequential implementation  in System I}
\label{fig:speedupI}
\end{figure}

The benchmarks can be clearly classified into two groups:
on Table \ref{tab:node-results} are the ones  where  cache-conscious decomposition brings considerable performance gains over horizontal decomposition, 
and on 
 Table \ref{tab:node-results2}, the ones where both decomposition strategies are on a par.
These results are clearly bound to the cache-locality properties of the benchmarks.
As could be expected, only the ones featuring both spatial and temporal locality  -  MatMult, MatTrans, GaussianBlur and SOR - really benefit from 
the cache-conscious decomposition.
The charts in Figures \ref{fig:speedupA} and  \ref{fig:speedupI}  
graphically substantiate this statement, presenting the speed-up  
 obtained by both decomposition approaches in the two systems.
The results are better for System A than for 
System I because, in the former, each core runs a single hardware thread, 
while, in the latter, cores are \textit{hyperthreaded} and thus run two hardware threads.
Consequently, in System A,  caches are shared by less threads and hence there are less conflict and capacity misses.

The performance gains for MatMult  and MatTrans are considerable (up to 6 - 7 times), specially as the size of the problem increases.
A major performance boost, since we are in presence of a data locality optimization not extra parallelization.
The results obtained for MatMult 1000 in System A  escape the norm, thus deserving a more careful explanation.
Given the size of the matrix, the number of workers, and the size of the L1 cache (64 KBytes),
the horizontal decomposition produces partitions 
not much bigger than the L1 cache.
Consequently, for this particular parameterization, the computation is already cache-friendly and there is not much room for cache-related optimizations.
The same does not apply to System I because the cache is smaller (32 Kbytes) and is shared by two hardware threads.
%
%Naturally, as the problem scales 

The gains for GaussianBlur and SOR are not as impressionable, but still very good: up to approximately 3 times better than the horizontal decomposition.
Note that these two benchmarks are stencil computations that, by operating over 2-dimensional neighbourhoods, already exhibit data access patterns  that leverages cache to some extension.
Naturally, as the size of the matrices and/or size the neighbourhood  increases, this property fades and the gains of 
cache-conscious decomposition are more noticeable.
Also note that  in GaussianBlur the computation is somehow unbalanced due to
the processing of the image's borders. This fact can  be further observed in the  speed-ups against the 
sequential version, which are relatively lower when compared with the remainder benchmarks. This is true for both decomposition strategies.

The benchmarks in the second set  do not benefit from temporal locality, and are thus  ideal for 
determining the overhead imposed by the cache-conscious decomposition, namely by
 the generation of a large
amount of tasks   and their subsequent   scheduling.
 In the  tested benchmarks no significant performance gains or losses can be found.
Crypt and Series iterate data sequentially (benefiting from spatial locality), without revisiting  previously accessed data.
Hence, no benefits are attained from enforcing temporal locality by keeping the partitions' size within the TCL.
Conversely,  
WordCount features temporal locality on the access to the map that stores the number of word occurrences.
However, the random access pattern to such data precludes any attempt to, in advance, predict which data to place in the TCL. 
A cache-aware implementation of such map is a challenge to overcome.
In sum, the presented  results show that the overhead of our approach
is negligible,  attesting that it may be 
used in a wide range of applications, without concerns for the type of locality.

\subsubsection{Sensitivity to the chosen TCL}
\begin{figure}
\centering
\includegraphics[width=0.8\linewidth ]{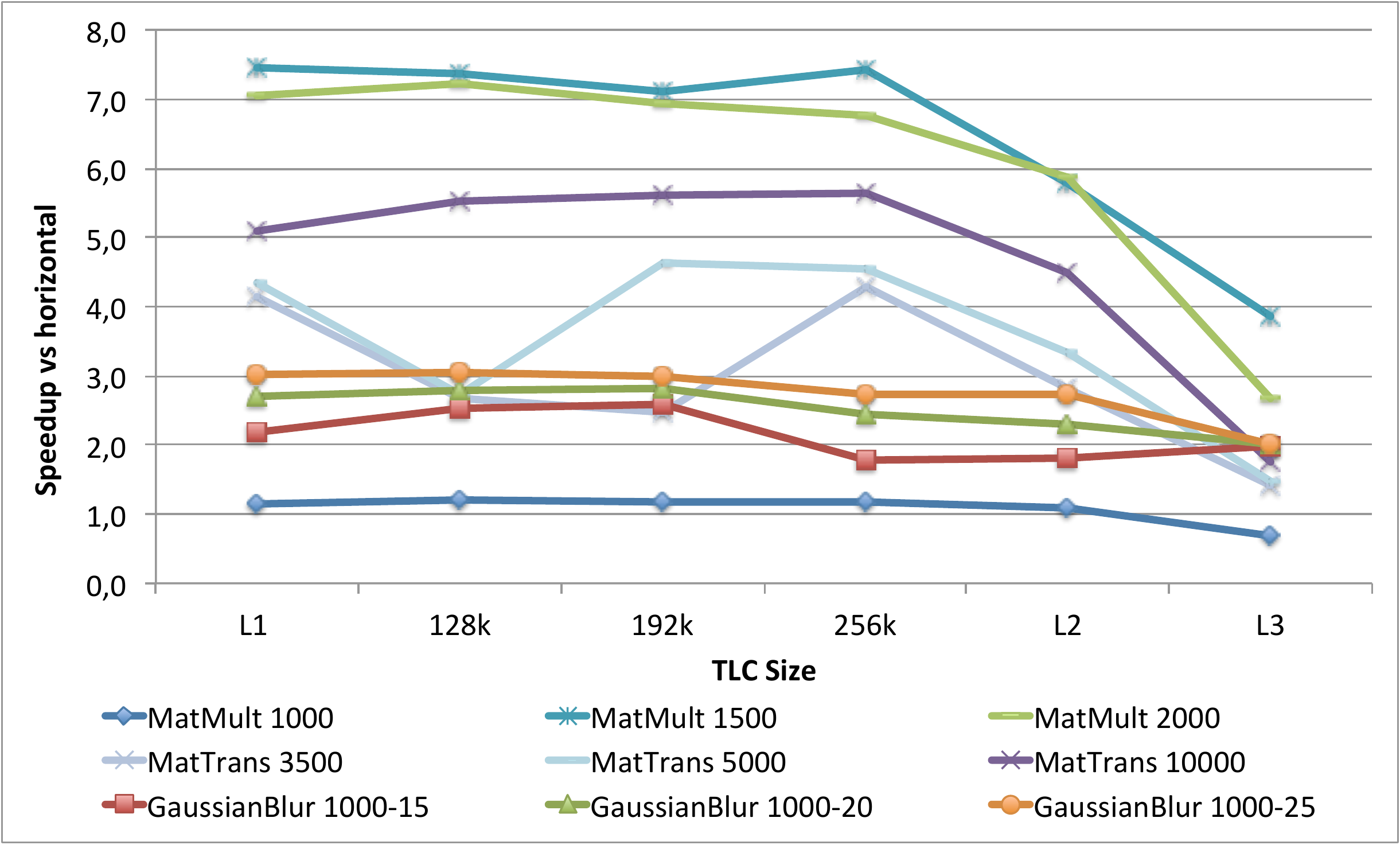}
\caption{Speed-ups of the CC strategy  in System A, varying the size of the TCL}
\label{fig:node1h}
\end{figure}

Next we perform a sensitivity analysis of  the results relatively to the chosen TCL
and its size.
%
%Our initial intuition was that for problems benefiting from data locality, the optimal size for a partition	 would be the size of the L1 data cache. However, some peculiar results drove us to experiment with other values.
%Consequently, in both systems, 
To that end we vary the size of the  TCL  from the L1 to the L3 cache sizes.
Given that the results are similar across all systems and scheduling strategies, we limit our discussion  to just one
configuration: CC scheduling strategy in System A (Figure \ref{fig:node1h}), and present the essence of the analysis in 	Table \ref{TCL}.
%The speed-ups (relatively to the horizontal decomposition) obtained for each considered TCL size 
%are depicted in
%Figure \ref{fig:node1h}.
%
The results are very insightful, as the optimal TCL value lies somewhere between the sizes of the L1 and the L2 caches and is
 benchmark and architecture dependent.
 We associate this to the fact that: (a) the JVM  itself has a state that competes for space in the cache, and 
 (b)  
 we are not considering all the dimensions of 
cache hardware, particularly the number of ways, and hence are permeable to conflict misses.

\begin{table*}
\centering
\includegraphics[width=\textwidth]{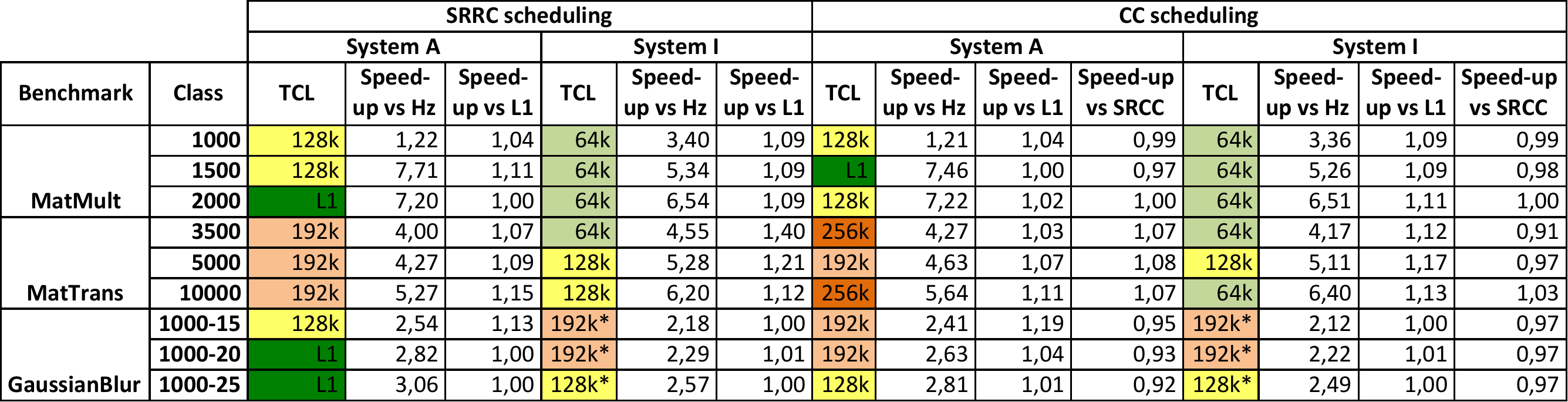}
\caption{Speed-ups of the CC and SRRC strategies varying the TCL size: summary table. 
The TCL values bearing the  * mark are
the smallest size  for which  valid partitioning solutions were found.}
\label{TCL}
\end{table*}

\subsubsection{Sensitivity to the scheduling strategies and to the $\varphi$ function}
Also from Table \ref{TCL} it can be observed that,  for most cases, the impact of employing the CC or the SRRC scheduling strategies
on the benchmark's absolute performance   is not substantial.
The  TCL for which the benchmark attains its performance peak does 
depend on the scheduling strategy employed, but the absolute performance of such peaks are very close - the essential  gains result from the base cache-conscious decomposition strategy. 
GaussianBlur is the sole benchmark to consistently benefit from one of the strategies, SRRC,  for higher radius values: larger blur window increases the amount of data that is shared by tasks operating over contiguous windows.

Regarding the $\varphi$ functions, we also evaluated the benchmarks using the two $\varphi$ functions 
presented in Section \ref{sec:decomp:node}.
The results showed that cache-line-awareness did 
not improve the performance for any of the assessed cache sizes.
In fact, the results were worse, given the introduced 
overhead and the wasted cache space of the conservative approach. 
%

%sensibilidade à precisão da função de particionamento 
%

\subsubsection{Breakdown}
\begin{figure}
\centering
\includegraphics[width=0.7\linewidth]{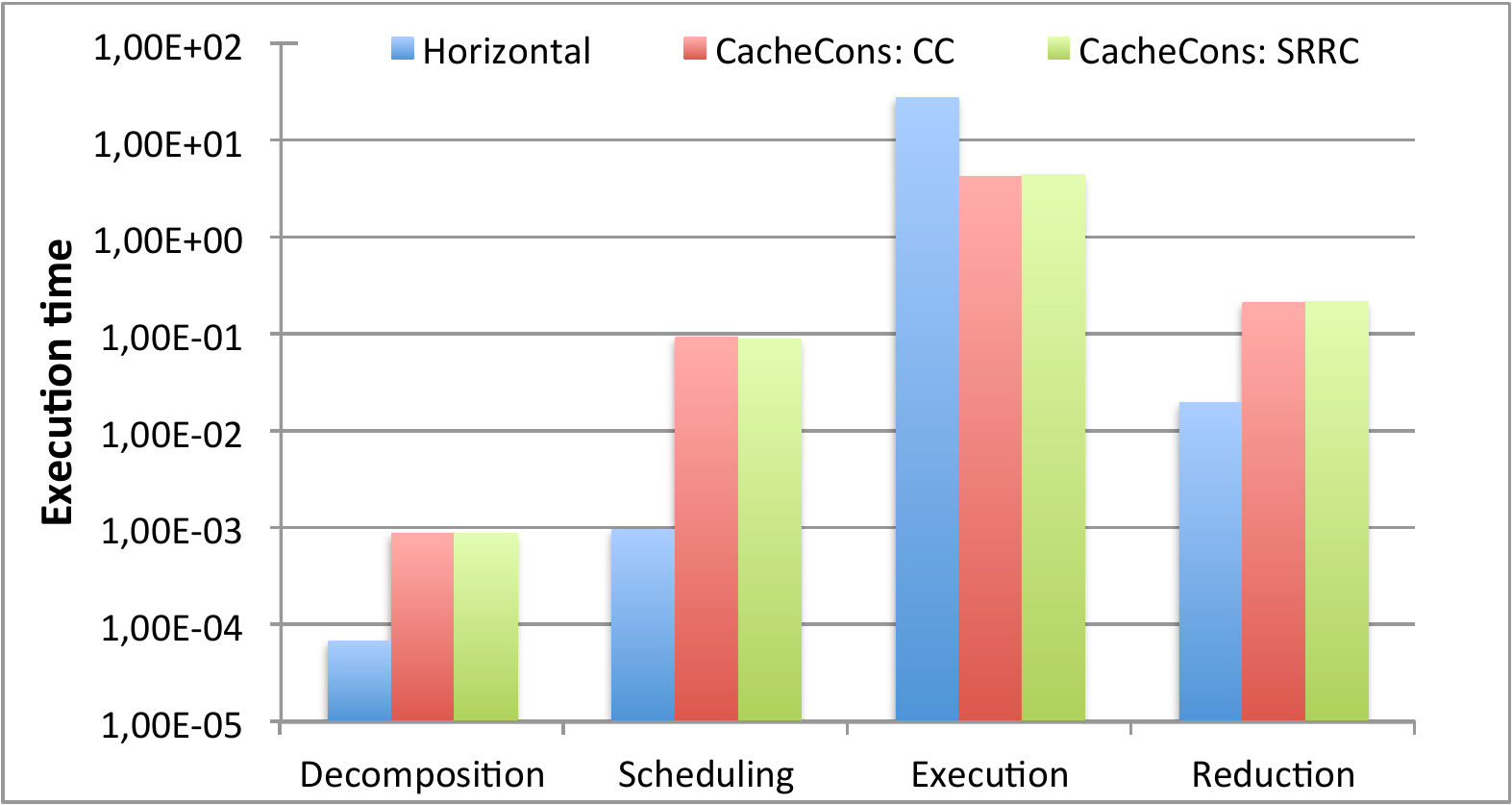}
\caption{Breakdown of MatMult N=2000 (System A) - logarithmic scale}
\label{fig:breakdown}
\end{figure}

In order to better evaluate the overhead imposed 
by the proposed  cache-conscious approach (denoted \textit{CacheCons} in Figure \ref{fig:breakdown}), 
we broke down the execution of the MatMult benchmark, in system A, for N=2000 and TCL size of 128k (the best performance).
\textsf{Decomposition} and \textsf{Scheduling} denote the time spent in the decomposition of the domain and the subsequent scheduling of the tasks,
\textsf{Execution} denotes the time spent in the actual matrix multiplication, and \textsf{Reduction} represents the time
spent in the reduction of the partial results (see Figure \ref{fig:matrix_combs}).

As depicted in the figure, the 
 weight of the  stages other that \texttt{Execution} is one order of magnitude higher in the cache-conscious  than 
in the horizontal approach.
This impact is mostly visible in the \textsf{Reduction} stage, that in the cache-conscious cases reaches almost 5\% of the whole execution time,
given that that number of results to reduce at node-level is much higher.
The remainder \textsf{Decomposition} and \textsf{Scheduling} stages are fully optimized for either clustering strategy, 
pertaining to less than 2\% of the total execution time.
In this particular example, the cache-conscious decomposition generates 8000 tasks (1000 per each of the 8 workers),
that are created and scheduled in less than a 0.1 seconds.

Despite the overhead imposed in the three aforesaid stages, the overall performance  of the cache-conscious is much better.
This is due to the substantial gains obtained in the heavier \textsf{Execution} stage, that  takes more than 
$90\%$ of the total execution time in all three cases.

\begin{figure}
\centering
\includegraphics[width=0.7\linewidth]{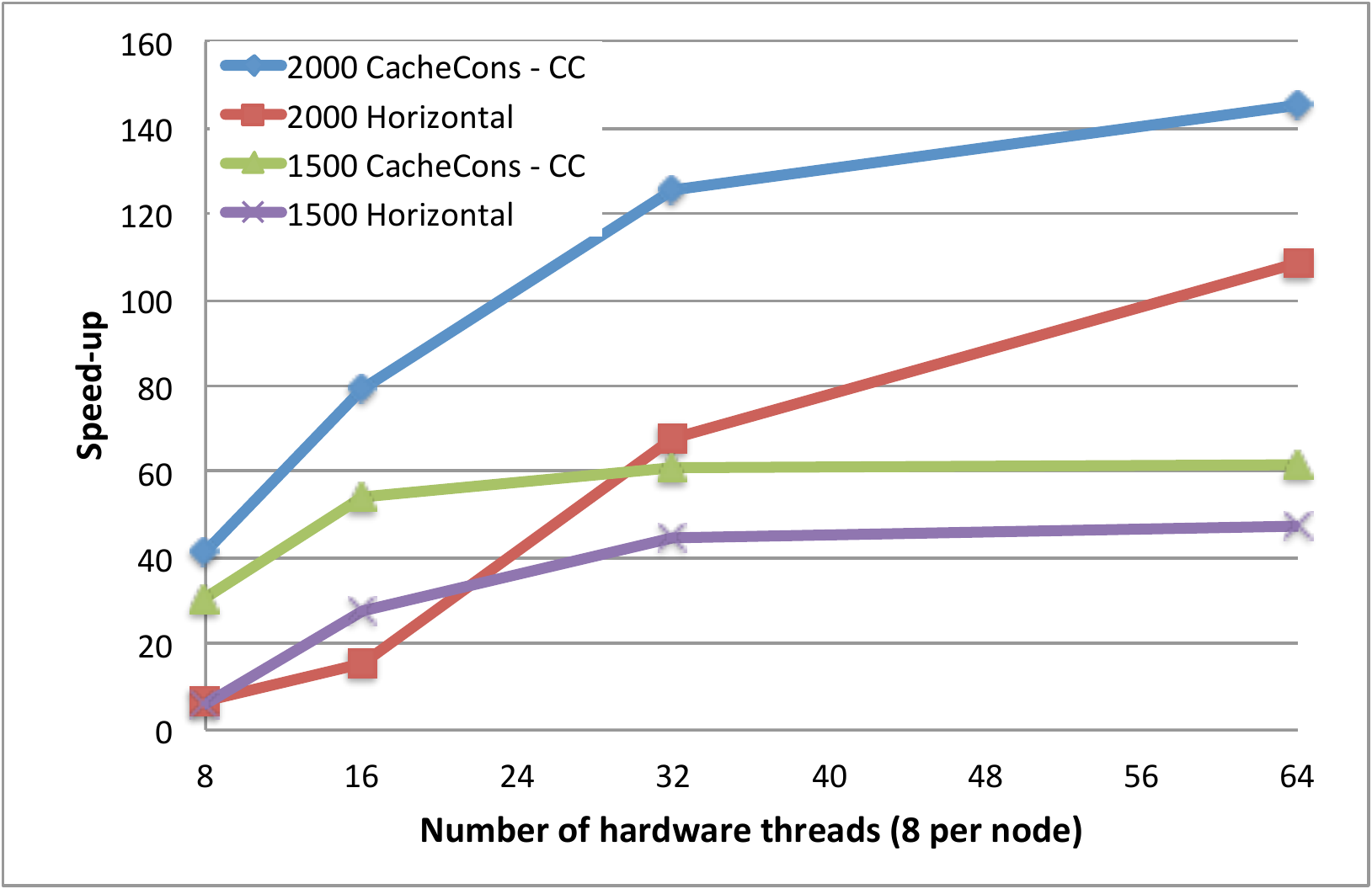}
\caption{Speed-up of the MatMult benchmark - Cluster environment}
\label{fig:cluster}
%\vspace{-15pt}
\end{figure}

\subsection{Impact at Cluster-Level}
To illustrate the impact of our approach at cluster-level, 
we present the speed-up of  two instances of the MatMult benchmark relatively to the sequential execution,
when varying the number  of nodes: 1, 2, 4 and 8  (each featuring 8 hardware threads).
Figure  \ref{fig:cluster} depicts the curves for the horizontal and cache-conscious decompositions.
For the 2000 problem class, both approaches scale up to 8 nodes (64 hardware threads).
Naturally, the node-level gains of the cache-conscious approach transpire in the overall execution time.
The performance peek of the cache-conscious decomposition 
delivers a 137 speed-up over the  sequential execution, and a 3.7
speed-up over  the horizontal approach.
In the horizontal approach, 
as the number of nodes increases, 
the data partition assigned to each node, and subsequently to each worker thread, diminishes
 and approximates itself to the size of the higher cache levels.
 This performance increase is most noticeable from the 2 to the 4 node configuration. 
As such, one can also leverage the impact of cache locality in  horizontal decompositions.
 However, these gains are ephemeral and  do not scale, giving that they are bound to the assignment of small data partitions to nodes.
Conversely, cache-conscious decomposition  delivers higher gains for data-sets of big dimensions.
In this sense, it is particularly suitable for  cluster environments, where the scale of the problems to solve
is, by definition, large.

\section{Related Work}
\label{sec:related} 

\subsection{Hierarchical Data Parallelism}

Our work is close to the hierarchical data parallelism field.
However, contrary to our proposal, 
hierarchical data parallelism models place on the programmer's shoulders the burden of decomposing the domain according to the traits of the memory hierarchy.
Of these models,  Sequoia \cite{sequoiapmh,sequoia_portable,Bauer_programmingthe} 
is the most prominent.
It provides a programming language 
for the explicit programming of the memory hierarchy.
The language's main program building block is the task,   a side-effect free function with call-by-value parameter passing semantics.
Through tasks the programmer is able to express: parallelism, explicit communication and locality, isolation, algorithmic variants and parameterization.
These properties allow programs written in Sequoia to be portable across machines without sacrificing the possibility of tuning the application for each one. %target machine.
Tasks run in isolation, and hence may be executed concurrently without requiring synchronization between cooperating threads. 
Parameter passing during task launching is the only communication mechanism available, %even  when tasks execute concurrently on the same machine, 
which increases the complexity of expressing cooperative computations.
Hierarchical awareness is achieved by providing different implementations (variants) of a given task. \emph{Inner} variants reflect intermediate nodes in the memory hierarchy and have the purpose of %generically
decomposing the input dataset.
\textit{leaf} variants perform the actual computation, operating directly on working sets residing within leaf levels of the memory hierarchy.
Mapping a hierarchy of tasks onto the hierarchical representation of memory requires the creation of task instances for every machine level involved.
The programmer is required to provide the compiler with the task mapping specification for the machine where the algorithm will be compiled and executed.
%

%No other mechanisms are provided for a task to communicate
%with other tasks that execute concurrently on the same machine.

Parallelism in Sequoia is assumed to be regular. 
Additional constructs to support irregular parallelism are proposed in \cite{Bauer_programmingthe}, namely the  \emph{call-up} and \emph{spawn}.  \emph{call-up}  allows subtasks to access their parent's heap, which can be used to modify its data structures. %in the latter.
%Since a task typically launches multiple subtasks, the \textsf{call-up} construct introduces concurrency into the Sequoia programming model.
%
\textit{spawn} provides for the dynamic  generation of parallelism, being able to launch an arbitrary number of subtasks of the provided task.
%%%%%%%%%%%

%%%%%%%%%%
%HTA
%%%%%%%%%%
Hierarchically Tiled Arrays (HTA) \cite{Biksh06programmingfor} is a programming paradigm 
that relies on a object type named \emph{tiled array} for expressing parallelism and locality.
HTAs are arrays partitioned into tiles, which can in turn be either conventional arrays or further tiled arrays, enabling hierarchical decomposition.
The components of an HTA can be accessed in a way analogous to the conventional array indexing. %although some additional indexing notation is added in order to make it easier for programmers to express some more complex accesses. % and to distinguish between scalar data and tiles accesses. 
%
%HTA programs are single-threaded, with  parallel computations being represented as array operations.
%Communication is expressed using assignments on distributed HTA or using other HTA operations, such as \textit{permute}, \textit{circshift} and \textit{repmat}.
%%%%%%%%%%%
Once more, the hierarchical decomposition is explicit in the program: the \texttt{level} function can be used to obtain, at runtime, the location of the argument within the tile hierarchy.
%CORTADO%
%This approach is equivalent to Sequoia's variants, since the programmer must incorporate hardware-awareness in its algorithm.

%%%%%%%%%%
%HPT
%%%%%%%%%%
The Hierarchical Place Trees (HPT) \cite{hpt} model combines concepts from Sequoia and the X10  languages.
From the latter it borrows the concepts of \textsf{place} and \textsf{activity} (task).
It abstracts each  memory module  as a place, and therefore a memory hierarchy is abstracted as a \emph{place tree}. Places are tagged with annotations that indicate their memory type and size. Moreover, a processor core is abstracted as a \emph{worker thread} which in the HPT model, can only be attached to leaf nodes in the place tree.
%The removal of this last restriction has been considered in order to accommodate \emph{processor-in-memory} hardware architectures in the HPT model.
%
In contrast with Sequoia, HPT supports three different types of communication: implicit access, explicit in-out parameters, and explicit asynchronous transfer. %In addition, HPT allows dynamic task scheduling, rather than static task assignment as in Sequoia.
%
%CORTADO%
%However, one more, hardware-awareness is entailed in the program.
%The  \texttt{Place} class features method \texttt{isLeafPlace} which is equivalent to HTA's\texttt{level}.

%%%%%%%%%%%
%Others
%%%%%%%%%%%
Less noticeable works are that also  address hierarchical decomposition are
\cite{hierarchical_upc} and \cite{Kamil:EECS-2013}. These systems perform horizontal decompositions 
at both the cluster and the node level,  firstly decomposing the domain at hand by the multiple nodes that compose the distributed environment, and secondly by the pool of workers available at each node.

There are several fundamental differences between our work and hierarchical data parallelism.
The most evident is (as previously stated) that we do propose a systematic approach to the memory-hierarchy-aware 
decomposition of data-sets, and not programming abstractions for the programmer to do so.
We do ask for some help from the programmer in the implementation of the \texttt{Distribution} interface, but it is architecture agnostic information.
Moreover, none of these models can be applied to the class of computations that we are aiming for: computations where the decomposition stage is cleanly separated from the computation stage. Hierarchical data parallel models follow a divide-and-conquer approach that blends  the two stages together.
Finally, in rigor, we do not follow a hierarchical approach, as we do not iteratively partition a domain so that it first fits the higher memory level, and from then on, each of the inferior levels.
%\
We delegate such enterprise of the scheduling strategy. 
For instance, the SRCC strategy tries to keep in each LLC the stream of partitions that will be feed to  workers running in the cores bound to that cache level.

\subsection{In-Memory MapReduce}

There has been quite some work on the optimization of the MapReduce execution model
to the intra-node reality.
These efforts, commonly referred to \textit{in-memory MapReduce}  show some concern about data locality. 
For instance, Phoenix \cite{DBLP:conf/hpca/RangerRPBK07} adjusts the size of the input and output data of a \emph{map} task, so that this data can fit in the L1 cache, which reveals a concern about the utilization of the cache. 
However, no attention is paid to the layout of the data in memory, nor to the overall organization of the memory hierarchy, namely
the core-to-cache affinities and the degree of sharing of cache levels among cores.
It is just an optional user-tuned parameter that conveys information about the preferred size for each partition.

Another system that presents a rationale close to ours is Tiled-MapReduce \cite{DBLP:journals/taco/ChenC13}.
It employs a pipeline of \textit{map} and \textit{reduce} tasks that make use of the same memory spaces and reduces idle time of the processing units, promoting locality. 
Furthermore, it also makes use of tiling strategies for the partition of the domain.
However, this  tilling process does not take the memory hierarchy into consideration, a key
contribution of our work.

Other data locality related concerns have been addressed in  in-memory MapReduce implementations.
Phoenix++ \cite{talbot2011phoenix++}  and Metis \cite{mao2010optimizing} have directed their
focus to the efficient implementation of the data structures that harbor the intermediate results of the \textit{map} stage.
Additionally, Phoenix++ also tries to take advantage of locality 
by applying the \textit{combiner} stage every time a new intermediate result  is emitted. The motivation is to, in an ad hoc manner, 
benefit from the likely possibility of the result still residing in a lower level cache. 

In turn, HJ-Hadoop \cite{DBLP:conf/oopsla/Zhang13} extends Hadoop with features of the Habanero Java framework for multi-core parallelism.
Data locality is explored by reducing  the number of Java virtual machines created per node, allowing for a more efficient  parallel execution of the \emph{map} stage and for the buffering of the data feed to the user-defined functions.

\section{Conclusions}
\label{sec:conclusions}
In this paper we defined and implemented a systematic cache-conscious  strategy for decomposing the domain of an application according to the traits of the target machine's memory hierarchy.
A performance evaluation demonstrates the advantage of the approach  when targeting computations that feature temporal locality in the access to data. We have obtained up to a 7.7  speed-up relatively to the standard horizontal decomposition in this class of computations.
In the remainder cases, no performance penalties were observed,  foreseeing a wide applicability of the solution.

Another important conclusion drown from this work is that the best 
clustering strategy, and TCL size configuration, is computation and architecture-dependent.
It is not possible to systematically use the same execution settings across applications and architecture,
which compromises performance portability.
To mitigate this problem, we are currently addressing the automatic inference of these configurations.
The goal is to design and implement an auto-learning stage that, over time, progressively learns the best configurations to be applied for each problem and its input sizes, applying these settings upon a request to execute the given problem.

Finally, we have also presented initial evidences that cache-conscious decomposition is also of particular usefulness in cluster environments, as it is improves the data locality of  algorithms that manipulate large data-sets.
Future work will assess its applicability to the area of Big Data analytics.
 %

%It may also be possible that  the optimal execution settings for each problem can be determined if more information about the memory  hierarchy is taken into consideration, examples of these being the cache-group-size and the cache-line removal heuristic.
%
%%This strategy was implemented on top of the Elina framework, whose modularity allowed us to test several decomposition and clustering strategies without modifying the applications' source code.
%%Accordingly, we presented, to the best of our knowledge,   the first performance comparison between horizontal and vertical decomposition strategies.
%%%
%%%This allowed us to conclude that vertical is applicable for the studied computations, which are MapReduce-like.
%%%
%%The obtained results favour our approach, with speed-ups peeking at 700\% for applications featuring temporal locality on the access to data.
%%
%%
%%\vspace{-5pt}

\bibliographystyle{elsarticle-num}
\bibliography{main}

% that's all folks
\end{document}